%% file: main.tex
\documentclass[acmsmall, nonacm]{acmart}

\newcommand{\BULLET}{\vspace{+.05in} \noindent $\bullet$ \hspace{+.00in}}

\newcommand{\keyrate}{\mathsf{R}}
\usepackage{mathtools}

\AtBeginDocument{%
  \providecommand\BibTeX{{%
    \normalfont B\kern-0.5em{\scshape i\kern-0.25em b}\kern-0.8em\TeX}}}

\usepackage{enumitem}
\usepackage{hyperref}
\usepackage{color,soul}
\usepackage{xcolor}   % added
\definecolor{purple}{RGB}{128,0,128} 

\input{bing-command}

\usepackage{stmaryrd}
\usepackage[skip = 0.15pt]{subcaption} %
\usepackage{multirow}
\usepackage{wrapfig}
\usepackage{array}
\usepackage{algorithm}
\usepackage{algorithmic}
\usepackage{amsmath}

\begin{document}

\iffalse
\setlength{\abovedisplayskip}{0.96pt}
\setlength{\belowdisplayskip}{0.96pt}
\fi

%\title{Optimal Scheduling for Single-downlink Satellite-based Quantum Key Distribution}

%wb, 9/25/2025, emphasize opportunistic 
\title{Opportunistic Scheduling for Single-downlink Satellite-based Quantum Key Distribution}

%%author list: Md Zakir Hossain, Nitish K. Panigrahy, Walter O. Krawec, Don Towsley, Bing Wang

% \author{Md Zakir Hossain}
% \author{Nitish K. Panigrahy}
% \author{Walter O. Krawec}
% \author{Don Towsley}
% \author{Bing Wang}
% \authornote{Corresponding author.}

\author{Md Zakir Hossain}
\affiliation{%
  \institution{University of Connecticut}
  \city{Storrs}
  \state{Connecticut}
  \country{USA}
}
% \email{zakir.hossain@uconn.edu}

\author{Nitish K. Panigrahy}
\affiliation{%
  \institution{Binghamton University}
  \city{Binghamton}
  \state{New York}
  \country{USA}
}
% \email{npanigrahy@binghamton.edu}

\author{Walter O. Krawec}
\affiliation{%
  \institution{University of Connecticut}
  \city{Storrs}
  \state{Connecticut}
  \country{USA}
}
% \email{walter.krawec@uconn.edu}

\author{Don Towsley}
\affiliation{%
  \institution{University of Massachusetts Amherst}
  \city{Amherst}
  \state{Massachusetts}
  \country{USA}
}
% \email{towsley@cs.umass.edu}

\author{Bing Wang}
\authornote{Corresponding author (bing@uconn.edu).}
\affiliation{%
  \institution{University of Connecticut}
  \city{Storrs}
  \state{Connecticut}
  \country{USA}
}
% \email{bing@uconn.edu}

%====================================================

%=========================================================
\input{abstract}

\maketitle

\input{intro}

\input{background}

\input{problem}

\input{opt-maxmin}

\input{framework}

%wb, 10/3/2025
% \input{sigmetrics/2-phase-methods}
\input{os}

\input{results}

\input{related}
\input{concl}

%wb, 9/25/2025, to decide later on
%\bibliographystyle{ACM-Reference-Format}
% \bibliographystyle{plain}
% \bibliography{bib/quantum,bib/quantum-1}

\bibliographystyle{ACM-Reference-Format}

\input{output}
\end{document}

%% file: bing-command.tex
\newcommand{\remove}[1]{}

%% file: abstract.tex
\begin{abstract}
Satellite-based quantum key distribution (QKD), leveraging low photon loss in free-space quantum communication, is widely regarded as one of the most promising directions to achieve global-scale QKD. With a constellation of satellites and a set of ground stations in a satellite-based QKD system, how to schedule satellites 
%to ground stations for efficient QKD  
to achieve efficient QKD is an important problem. This problem has been studied in the dual-downlink architecture, where each satellite distributes pairs of entanglements to two ground stations simultaneously. However, it has not been studied in the single downlink architecture, where satellites create keys with each individual ground station, and then serve as trusted nodes to create keys between pairs of ground stations. While the single downlink architecture provides weaker security in that the satellites need to be trusted, it has many advantages, including the potential of achieving significantly higher key rates, and generating keys between pairs of ground stations that are far away from each other and cannot be served using the dual-downlink architecture. In this paper, we propose a novel opportunistic approach for satellite scheduling that accounts for fairness among the ground station pairs, while taking advantage of the dynamic satellite channels to maximize the system performance. We evaluate this approach in a wide range of settings and demonstrate that it provides the best tradeoffs in terms of 
total and minimum key rates  across the ground station pairs. Our evaluation also highlights the importance of considering seasonal effects and cloud coverage in evaluating satellite-based QKD systems. In addition, we show different tradeoffs in global and regional QKD systems. 
\end{abstract}

%several approaches for scheduling satellites in the single downlink architecture with the goal of maximizing the total key rate, or maximizing the minimum key rate, over all ground station pairs. We evaluate these approaches in a wide range of settings and present tradeoffs in total key rate and fairness. 

%% file: intro.tex
\iffalse 
1. Why do we consider single downlink, not dual (coverage issue, maybe we can show a plot)?
2. Why is scheduling critical, that to (max-min based) fairness? 
3. Why is single downlink scheduling a difficult problem (single phase vs two phase)? We need to have some auxiliary plots (like execution time vs problem size for max-min).
4. In the background, we can add some maths related to single downlink QKD. Even though some of the concepts are well known, I am coming across several papers that revisit known ideas, while providing plots and explanations to reinforce the importance of the problem they address.
\fi 

\section{Introduction}

%directly talk about satellite-based QKD; two ground stations $A$ and $B$ established shared secret keys leveraging QKD protocols, which provides information-theoretic security. 
%mode of operation: dual downlink; single downlink; why downlink (All three modes use downlink transmission, which has been shown to be advantageous compared to uplink transmission (i.e., from ground to the satellite) [57]); advantages of both approaches
%

%In the dual-downlink mode, a satellite sends entangled pairs to two ground stations simultaneously to establish shared secret key directly between them. In the single-downlink mode, a satellite sends photons to ground stations individually, and then serves as a common node to establish shared keys between two ground stations.

%Compared to the dual-downlink mode, the single-downlink mode requires fewer and simpler resources, and has a higher success rate. However, unlike the dual-downlink mode, it requires the satellites to be trusted, and its classical post-processing of QKD involves both the satellites and ground stations.

%why opportunistic scheduling? Why it is suitable here? What are the main advantages?

%10/1/2025, novel

Satellite-based quantum key distribution (QKD) 
allows two ground stations $A$ and $B$ that are far away from each other to establish shared secret keys using QKD protocols that provide information-theoretic security~\cite{Pirandola19:QKD-Survey}.
%, with no computation assumptions~\cite{}. 
It is widely regarded as one of the most promising directions to achieve global-scale QKD, since free-space quantum communication between  satellites and ground stations leads to much lower photon loss compared to photon communication on the ground~\cite{Khatri21:Spooky,Simon17:Towards,Aspelmeyer03:satellites,Jennewein13:space-race,Bedington17:satellite,Bae2025:Blockwise}. 
The feasibility of satellite-based QKD has been demonstrated experimentally  \cite{wang2013direct,yin2013experimental,vallone2015experimental,liao2017satellite}. On the other hand, many challenges remain to make satellite-based QKD into an efficient and commercially-viable system. 

%The advantage of QKD compared to classical key establishment protocols lies in that QKD  provides information-theoretic security, with no computation assumptions~\cite{}.  

%Consider a constellation of satellites that assists secret key establishment among a set of ground stations. 
Consider a satellite-based QKD system with a constellation of satellites and a set of ground stations. The quantum communication can be {\em downlink}, from satellites to ground stations, or {\em uplink}, from ground stations to satellites. The downlink direction has an advantage over the uplink direction due to lower photon loss in the initial stage~\cite{Bedington17:satellite}. In this paper, we focus on a {\em single-downlink} architecture, where a satellite sends photons to ground stations individually, and then serves as a common node to establish shared keys between two ground stations. This differs from a dual-downlink architecture, where a satellite sends entangled pairs to two ground stations {\em simultaneously} to establish a shared secret key directly between them. 
While single-downlink architecture requires that satellites need to be trusted (since they serve as trusted nodes), and hence has weaker security than the dual-downlink architecture where the satellites do not need to be trusted, it has several other advantages in terms of cost, efficiency and coverage. First, the single-downlink architecture only requires satellites to generate photons, instead of entanglements, and hence are simpler and less costly than the dual-downlink architecture. Secondly, since the single-downlink architecture involves quantum communication along one link, i.e., between a satellite and a ground station, instead of along two links, between a satellite and two ground station simultaneously, it has a higher success rate per round, and hence can produce a higher key rate than the dual-downlink architecture. Last, the single-downlink architecture       
does not require a satellite to be in view of   
two ground stations simultaneously, hence it can achieve key establishment among ground stations that are much farther apart than what is feasible with dual-downlink settings. As an example, in a prior study~\cite{Maule2024:scheduling}, no key could be established between New York City and Houston in the U.S. when using satellites at altitude of 500 km (polar constellation) in the dual-downlink setting, while   it can be easily achieved in the single-donwlink setting (see \S\ref{sec:eval}). 

Since a ground station can be served by multiple satellites in the single-downlink architecture at one point of time, satellite scheduling determines which satellites serve which ground stations.
%at one point of time. 
A scheduling algorithm needs to satisfy the constraints of the satellites and ground stations. It also needs to be efficient, maximizing system performance, while being fair to the ground stations. While satellite scheduling has been studied in the dual-downlink setting~\cite{panigrahy2022optimal, williams2024scalable, chang2023entanglement, wei2024optimizing,Maule2024:scheduling}, scheduling in the single downlink architecture for key establishment among ground station pairs has not received much attention  (see \S\ref{sec:related}). In this paper, we formulate the problem and develop efficient solutions that consider both system performance and fairness among the ground station pairs.

Our work makes the following contributions: 

\BULLET We formulate the satellite scheduling problem for the single-downlink setting for QKD systems and develop two comparison baselines that optimize minimum and total key sizes across the ground station pairs, respectively (\S\ref{sec:problem}). 

\BULLET We develop an opportunistic scheduling framework where schedulers take advantage of the dynamic and diverse satellite to ground station channels for efficient key establishment among ground station pairs 
%assisted by the satellites 
(\S\ref{sec:os-framework}). Our approach works in two phases. Phase 1 extends  opportunistic scheduling approaches~\cite{Liu2001:oppor,Liu2003:oppor} in classical wireless communication to multi-satellite settings for key establishment between satellites and ground stations (\S\ref{sec:sat-scheduling}). Phase 2 uses an iterative optimization approach to establish keys for ground station pairs assisted by the satellites. This two-phase approach can balance fairness and total key size, and has significantly lower computation overhead than the two baseline schemes. 

\BULLET Using extensive simulation in a wide range of settings, we demonstrate that our opportunistic scheduling framework achieves the best tradeoffs in terms of total and minimum key rates across the ground station pairs among all the schemes we evaluate (\S\ref{sec:eval}). Our evaluation further highlights the importance of considering seasonal effects and cloud coverage in evaluating satellite-based QKD systems. In addition, we show different tradeoffs in global and regional QKD systems.

%% file: background.tex
%\section{Problem setting and background}

\section{Background}
\subsection{Single-downlink Satellite-assisted QKD} \label{sec:sate-QKD}

\begin{figure}[t]
  \centering
  \includegraphics[width=0.5\linewidth]{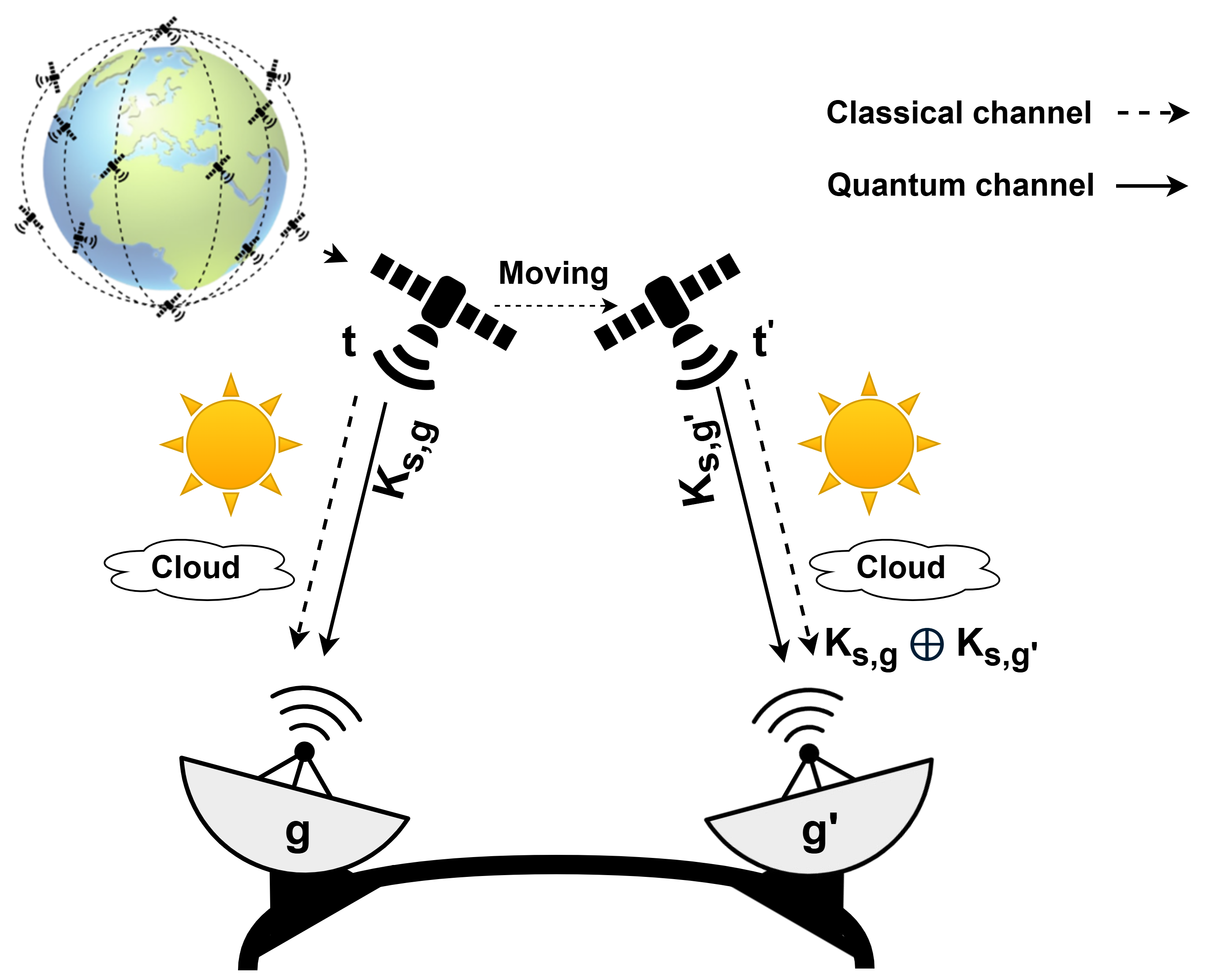} % or .png/.jpg
  \vspace{-0.1in}
  \caption{Single-downlink satellite-assisted QKD system. A satellite runs a QKD protocol with individual ground stations, and then uses classical channels and one-time pad to establish secret keys for ground station pairs.}
  \label{fig:single-downlink}
\end{figure}

We consider a constellation of satellites in low-earth orbit (LEO) that facilitate QKD for a set of ground stations on Earth; see Fig.~\ref{fig:single-downlink}.  In the rest of the paper, we assume that the satellites follow a Polar constellation; our approach is applicable to other types of constellations (e.g., Walker, Iridium, Starlink, and Kuiper~\cite{Khatri21:Spooky}). Similarly, our approach is not limited to LEO satellites; we focus on LEO satellites due to their proximity to the Earth, which can lead to high key rates.

In the single-downlink setting, a satellite establishes secret keys with each ground station individually using a prepare-and-measure QKD protocol, e.g., BB84~\cite{BB14:QKD}. Specifically, each satellite is equipped with a photon generation source.
%and transmits photons to one ground station in view.  
When a satellite is in view of a ground station, it generates  quantum states and sends them %quantum state 
to the ground station through a free-space optical channel. The ground station measures the received photons using randomly chosen bases, resulting in correlated raw key bits between the two parties.

%\wok{Changed things below, Bing can you please check when you have time?} 
We next briefly describe the BB84 QKD protocol, which we use in the rest of the paper. For a general QKD survey, see \cite{pirandola2020advances}.  QKD protocols operate in two stages: a quantum communication stage, followed by a classical post processing stage.  During the former, photons (which can encode quantum bits through polarization encoding \cite{pirandola2020advances}) are transmitted from the satellite, which attempts to encode secret key material onto the photon by encoding the classical data using one of two mutually unbiased bases (MUBs), chosen randomly. Two bases are mutually unbiased if the magnitude squared of the inner product of any vector from one, with any vector from another, is $1/d$, with $d$ being the dimension of the system; $d=2$ in the qubit case.  These photons/qubits travel to the ground station where they are subsequently measured, with the measurement being performed in one of two MUBs. If the satellite and the ground station select the same basis, then the satellite and ground station should have a correlated bit; otherwise, the round is discarded. Over many rounds, the satellite and the ground station obtain correlated bits and pool them into a \emph{raw key}.  These raw keys (one held by the satellite and one by the ground station) are partially correlated, as noise may have caused some measurement error, and partially secret, an adversary may have some partial information on the raw key.  Thus, they cannot be used directly as secret keys and must
%, instead, 
be further processed through additional classical processing in the second stage.

The post-processing phase utilizes a classical authenticated channel, and consists of two steps. First, an error-correction protocol is applied to reconcile discrepancies caused by channel losses, background photons, and detector imperfections. Second, privacy amplification is performed which takes in the, now error corrected, raw key and outputs a new, smaller, secret key.  This is done by mapping the raw key through a two-universal hash function. The outcome is a shorter, but secure, secret key shared between the satellite and the ground station which is subsequently added to a \emph{key-pool} for these users (i.e., a secret bit string, held by both parties, that is both secret and fully correlated).  Note that this key-pool may also be considered a single, large, secret key.

An important question is: given the observed noise level in the raw key, how large will the final secret key be, after privacy amplification is run?    Let $N$ denote the number of raw key rounds exchanged between satellite $s$ and ground station $g$.  After error correction and privacy amplification, let $\ell$ denote the size of the final secret key (after privacy amplification). The \emph{secret key rate} is, then, defined as $r = \ell/N$.

%the following ratio:
%\begin{equation}
%r = \frac{\ell}{N}.
%\end{equation}

%wb, 10/11/2025, use E to represent error rate since Q is used to represent policy later on
In the asymptotic regime, where $N \to \infty$, the secret key rate can be expressed as \cite{renner2008security}
\begin{equation}
r = 1 - 2h(E), \label{eq:key-rate}
\end{equation}
where $E$ is the error rate of the raw key, and
$h(x) = -x \log x - (1-x)\log(1-x)$
is the binary entropy function.  Note that, in practice, $E$ can be estimated through classical sampling methods, by having the satellite and ground station disclose a random portion of their produced raw key, and subsequently discarding the sampled portion of the key.

Of course, the above is for one satellite and one ground station.  However, the goal is to allow two ground stations to establish a shared secret key.  For this, the satellite will be used as a trusted node in the following manner. Let $K_{s,g}$ represent the set of key bits established between satellite $s$ and ground station $g$ after a sufficient number of rounds of QKD (i.e., $K_{s,g}$ is the key-pool for this particular satellite and ground station pair). Similarly, let $K_{s,g'}$, represent the key-pool shared between satellite $s$ and a different ground station $g'$. Then, through classical communication channels, satellite $s$ sends $K_{s,g} \oplus K_{s,g'}$ to $g'$. This allows $g'$ to learn $K_{s,g}$ by performing the operation $K_{s,g} \oplus K_{s,g'} \oplus K_{s,g'}$.  At this point, $K_{s,g}$ is now a shared secret key between ground station $g$ and $g'$.

In the above process, the secret key pool  $K_{s,g'}$, shared, initially, between $s$ and $g'$, was used as a secret key to encrypt $K_{s,g}$ via the one-time pad encryption algorithm, which is a perfectly secret method of encryption \cite{katz2007introduction}.  However, OTP requires that a secret key be used only once and that the key be as large as the message being encrypted.  This puts two constraints on the above process: First, once $K_{s,g'}$ is used to allow $g$ and $g'$ to share a key (namely, to share $K_{s,g}$), then it cannot be used again and must be discarded.  Second, $K_{s,g'}$ must be as large as, or larger than, $K_{s,g}$.  If it is shorter, then only $|K_{s,g'}|$ bits may be encrypted.  Thus, ultimately, $g$ and $g'$ will have a secret key of size $\min(|K_{s,g}|, |K_{s,g'}|)$; the rest of the key pool can be used later.

%By repeating this process across different time slots, a satellite generates secret keys with multiple ground stations. These independently established keys can then be combined through classical operations at the satellite, enabling ground stations to share end-to-end secure keys.
\subsection{Loss and Noise Models} \label{sec:model}

%\bing{Nitish, can you go over this subsection?} 
The free-space optical (FSO) channel between a satellite and a ground station introduces loss and noise that strongly influence the performance of quantum key distribution. Transmission loss increases quadratically with propagation distance and is further affected by the elevation angle\footnote{The elevation angle, $\theta_{s,g}(t)$, is defined as the angle between the horizontal plane at $g$ and the line-of-sight from $g$ to the satellite $s$.} and atmospheric conditions \cite{Maule2024:scheduling}.

Consider a satellite $s$ and ground station $g.$ At time $t,$ let the elevation angle be $\theta_{s,g}(t)$ and the line-of-sight distance between the satellite and ground station be $D_{s,g}(t).$ Let $\alpha_g(t)$ denote atmospheric profile above ground station $g$ at time $t$. This profile captures time-varying environmental parameters, including temperature, pressure and humidity, that affect photon attenuation in the atmosphere.
%The seasonal atmospheric profile is captured in $\alpha$.
%$\alpha\in \{\text{March},\text{ June},\text{ September},\text{ December}\}.$
In particular, for each ground station, we consider four representative atmospheric profiles corresponding to  the solstices and equinoxes (see later). We model the optical channel as a Bosonic pure-loss channel with time-varying transmissivity $\eta_{s,g}(t)$, which attenuates the transmitted quantum state and reduces the probability of successful photon detection. %In this model, a pure input state becomes mixed under non-zero loss, which in turn increases the quantum bit error rate (QBER). 
We compute the end-to-end transmissivity as:
\begin{align}
    \eta_{s,g}(t) = \underbrace{\eta_{\text{fs}}(D_{s,g}(t))}_{\text{free-space}}\cdot\underbrace{\eta_{\text{atm}}(\alpha_g(t), \theta_{s,g}(t))}_{\text{atmosphere}}\cdot \underbrace{\eta_{opt}}_{\text{optics}}.
\end{align}
\noindent Here, $\eta_{opt}$ captures coupling and other internal losses at the source and receiver hardware. The free space transmissivity $\eta_{\text{fs}}$ captures photon losses due to diffraction and decays quadratically as a function of $D_{s,g}(t).$ The atmospheric transmissivity $\eta_{\text{atm}}$ accounts for  photon losses due to absorption and scattering in the atmosphere and is expressed as \cite{dequal2021feasibility}:
\begin{equation}
\eta_{\text{atm}}(\alpha_g(t), \theta_{s,g}(t)) = [\hat{\eta}_{\text{atm}}(\alpha_g(t))]^{\text{cosec}(\theta_{s,g}(t))},
\end{equation}
\noindent where $\hat{\eta}_{\text{atm}}$ is the atmospheric transmissivity at zenith. Seasonal variation is incorporated by generating atmospheric profiles for representative days in March, June, September, and December at each ground station location, which are used in our evaluation in \S\ref{sec:eval}. To compute $\hat{\eta}_{\text{atm}}$ for a given atmospheric profile $\alpha_g(t)$, we use MODTRAN (Moderate Spectral Resolution Radiative Transfer Model) software \cite{MODTRAN} under clear-sky conditions with complete visibility and zero cloud cover.

Clouds introduce losses or in some cases complete blockage of satellite-to-ground optical links. To model this effect, we incorporate time and location dependent cloud coverage data, denoted by a cloud factor $c_{t,g} \in [0,1]$, where $c_{t,g}=0$ corresponds to clear sky and $c_{t,g}=1$ represents full obstruction of the downlink at ground station $g$ at time $t$. We obtain these cloud coverage data at an hourly scale from the Visual Crossing Weather API \cite{visualcrossing}.

In addition to transmission losses, quantum states encoded in photons are impaired by background photons and detector dark counts. During daylight hours, these background photons primarily originate from ambient solar radiation  and contributes to unwanted detection events at the ground station. The background photon count is highly time-dependent: during daylight hours, especially near noon, solar radiation induces elevated noise levels, whereas at night, the background photon flux is substantially lower. We evaluate this variation by sampling background photon flux at four representative times: 12:00 AM, 6:00 AM, 12:00 PM and 6:00 PM, for each ground station location. We model the arrival of background photons at the ground station telescope as spurious detection events that increase the quantum bit error rate.
%QBER. 
Finally, detector dark counts are included as a constant probability of false clicks per detection window. Together, these effects determine the overall signal photon detection probability, the quantum bit error rate, and ultimately the achievable secret key rate in single-downlink QKD.

%% file: problem.tex
\section{Problem Setting and Comparison Baselines} \label{sec:problem}

In this section, we first present the problem setting, and then two optimization-based formulations  that will be used as comparison baselines to evaluate our proposed opportunistic solutions.
%(see \S\ref{sec:os-framework} and ). 

\subsection{Problem Setting}
Let $\mathcal{S}$ denote a set of satellites and  $\mathcal{G}$ denote a set of ground stations. Consider a time interval (e.g., a day). 
Time is divided into discrete slots (e.g., each slot is one second) indexed by $t \in \{1,2,\ldots,T\}$. During each slot, a satellite $s \in \mathcal{S}$ may run QKD  with a ground station $g \in \mathcal{G}$ using the downlink quantum channel, provided that the satellite is above the horizon of the ground station and satisfies a minimum elevation angle requirement $\theta$.

%We consider the problem of scheduling satellites to ground stations in a time period, $T$ (e.g., a day). 
%The basic time unit for scheduling is a time slot (e.g., 1 second). 
At any point of time, a satellite can be in view of multiple ground stations. Similarly, a ground station can be in view of multiple satellites. 
%An important question is determining which ground station that the satellite sends photons to perform QKD. 
Let $M_s \ge 1$ denote the number of transmitters at satellite $s$ and $R_g \ge 1$ denote the number of receivers at ground station $g$. In other words, in each slot, satellite $s$ can serve up to $M_s$ ground stations, while ground station $g$ can be served by up to $R_g$ satellites. The scheduling problem is to determine, for any slot $t$, a subset of satellites (no more than $R_g$) in $\mathcal{S}$ that will serve each ground station, $g \in \mathcal{G}$ so that  the constraints of each satellite and ground station are satisfied. The goal is to achieve a high aggregate key rate across all satellite ground station assignments
%resource utilization of the system is high,
while providing a fair allocation of resources among ground station pairs.

For slot $t$, let $\lambda_t^{s,g}$ denote the number of photons that satellite $s$ distributes to ground station $g$ successfully in the slot, and let  $\keyrate_t^{s,g}$ denote the corresponding key-rate. Prior to computing a schedule, we first estimate $\lambda_t^{s,g}$ using the loss model in \S\ref{sec:model}. Similarly, we estimate $\keyrate_t^{s,g}$ by accounting for errors using the noise model in  \S\ref{sec:model} and key rate expression (\ref{eq:key-rate}). 

Let $c_{t,g}$ denote the cloud coverage for ground station $g$ in slot $t$, which is also estimated ahead of time based on weather prediction as described in \S\ref{sec:model}. 
 Therefore, for slot $t$, following the linear approximation in~\cite{Polnik20:scheduling}, the number of secret keys generated by satellite $s$ serving ground station  $g$ is $(1-c_{t,g})\lambda_{t}^{s,g} \keyrate_t^{s,g}$. In the rest of the paper, for ease of exposition, let $n_{t}^{s,g} \coloneq (1-c_{t,g})\lambda_{t}^{s,g} \keyrate_t^{s,g}$ denote the number of secret keys that can be generated in slot $t$ between satellite $s$ and ground station $g$.

%Consider a pair of ground stations $(g,g')$. In single downlink scenario, unlike the dual-downlink case, a satellite does not need to serve these two ground stations simultaneously. As a result, two ground stations that are far away from each other, and hence cannot be served by one common satellite simultaneously at any point, can be served in single downlink scenario. In contrast, in dual-downlink scenario, two ground stations need to be served by one satellite simultaneously. For example, no key can be created between NYC and London in dual-downlink scenario, while keys can created for them in the single-downlink scenario. 

%In the following, we first present a two-phase heuristic approach, and then present a single-phase optimization approach. 

Following a scheduling algorithm, let $K_{s,g}$ represent the set of key bits that satellite $s$ has established with ground station $g$ at the end of time $T$.
%the time interval. 
Since the goal is establishing shared keys among ground station pairs, satellite $s$ needs to further determine what fraction of the key bits in $K_{s,g}$ is used to create shared keys with another ground station, $g'$. Let  $\mathcal{U}=\{u \mid u=(g,g'), g' > g\}$ denote the set of ground station pairs where, for each ground station pair $u=(g,g')$, we represent the ground stations in increasing order of their indices. 
%Let $u=(g,g')$ represent the pair of ground stations $g$ and $g'$. 
%Then satellite $s$ needs to determine $y_{s,u}$, which represent the number of keys that satellite $s$ allocates to ground station pair $u$. 
Let $y_{s,u}$ represent the number of key bits that are established for ground station pair $u$ through satellite $s$. 
Since the shared key bits for $u$ via $s$ are obtained by XOR'ing the key bits in $K_{s,g}$ with those in $K_{s,g'}$, we need $|K_{s,g}| \ge y_{s,u}$, and $|K_{s,g'}|\ge y_{s,u}$, where $|K_{s,g}|$ represents the number of key bits in $K_{s,g}$.
The size of the key 
%total number of key bits 
that is established for ground station pair $u$ considering all the satellites is therefore $\sum_{s} y_{s,u}$. 

One goal of satellite scheduling is to achieve maxmin optimization, i.e., maximizing the minimum key size 
%of $Y_{s,u}$
%$\sum_{s} y_{s,u}$ 
across all the ground station pairs. However, since some ground stations may be in unfavorable conditions (e.g., due to its position or adverse weather conditions), such a scheduling  objective may result in a reduced total key size across all ground station pairs.
Another optimization goal is to maximize the total key size, which, however, may cause some ground station pairs to have small key sizes. Therefore, a good goal is balancing both fairness and total key size.

%wb, 10/1/2025
\iffalse 
Let  $\mathcal{U}=\{u \mid u=(g,g'), g' > g\}$ denote the set of ground station pairs, where we represent ground station pairs in increasing order of their indices. Let $y_{s,u}$ be the number of keys that satellite $s$ allocates to ground station pair $u$. Then the total number of keys for $u$ is $\sum_{s} y_{s,u}$. Consider a ground station pair, $u=(g,g')$. Let $K_{s,g}$ represents the number of keys that satellite $s$ has established with ground station $g$. Then we need $K_{s,g} \ge y_{s,u}$, and $K_{s,g}\ge y_{s,u}$. 

We develop two framework for determining the number of keys  between a pair of ground stations $(g,g') \in \mathcal{U}$. The first framework directly determines $y_{s,u}$ for all $u \in \mathcal{U}$. The second framework divides the problem into two phases, the first phase determines $K_{s,g}$, $\forall s  \in \mathcal{S}$ and $g \in \mathcal{G}$, and the second phase further determines $y_{s,u}$.  Henceforth, we refer to these two frameworks as {\em Single-phase} and {\em Two-phase} approaches.

We say that a ground station $g \in u$ if $g$ is one ground station in the ground station pair $u$. As an example, for three satellites $\mathcal{G}=\{1,2,3\}$, $\mathcal{U}=\{(1,2), (1,3), (2,3)\}$, ground station $1$ is in $(1,2)$ and $(1,3)$, ground station $2$ is in $(1,2)$ and $(2,3)$, and ground station $3$ is in $(1,3)$ and $(2,3)$. In $\mathcal{U}$, we only represent ground station pairs in increasing order of their indices (i.e., we only say $(1,2)$; no need to say $(2,1)$). 
\fi

%% file: opt-maxmin.tex
%wb, 10/1/2025
\vspace{-0.2in}
\subsection{Comparison Baselines} \label{sec:opt}
%: Optimization-based Solutions}
We next present two optimization formulations for satellite scheduling, {\em Max-Min} and {\em Max-Sum}. Max-Min aims to maximize the minimum of $\sum_s y_{s,u}$. Max-Sum aims to maximize $\sum_{s}\sum_{u} y_{s,u}$, i.e., the total key size across all the ground station pairs through all the satellites. Let $K_{\text{maxmin}}$ and $K_{\text{maxsum}}$ denote respectively the objective values from these two optimization problems. Then a good scheduling algorithm should have the minimum key size across all the ground station pairs close to  $K_{\text{maxmin}}$ and the total key size across all the ground station pairs close to $K_{\text{maxsum}}$. In the rest of the paper, we use these two quantities to evaluate satellite scheduling algorithms. 

%wb, 10/4/2025, change all t to be superscript to be consistent with later on; maybe not
Consider slot $t$. Let 
%$x_{s,g}^t$
$x_t^{s,g}$ 
denote a binary decision variable; $x_t^{s,g}=1$ if 
%that determines whether 
satellite $s$ serves ground station $g$ in slot $t$, and $x_t^{s,g}=0$ otherwise. Another decision variable is integer variable, $y_{s,u}$, which represents the size of the key  established for ground station pair $u$ through satellite $s$. The two optimization problems consider all the slots in $\{1,\ldots,T\}$
as follows
%It is an integer value that is to be determined in the optimization formulation. 

\iffalse 
\section{Single-phase Optimization}
Recall that $\mathcal{U}=\{u \mid u=(g,g'), g' > g\}$ denotes the set of ground station pairs. We next first present a formulation that considers the entire time interval $[1,T]$. This formulation involves a large number of decision variables for a large number of ground stations, and may not be solvable directly. 
%We therefore also present a window-based formulation that divides solves $[1,T]$ into multiple windows, and solve the optimization problem for each window sequentially.

\subsection{Maxmin Formulation}
The decision variables are $x_t^{s,g}$ and $y_{s,u}$, $\forall s \in \mathcal{S}$, $g \in \mathcal{G}$, and $u \in \mathcal{U}$.
\fi 

\begin{align}
\text{Max-Min:}\quad \text{maximize:} & \quad \min_u {\sum_{s \in \mathcal{S}}y_{s,u}} \label{eq:goal-maxmin-single-downlink} \\
\text{Max-Sum:}\quad \text{maximize:} & \quad \sum_{u \in \mathcal{U}}{\sum_{s \in \mathcal{S}}y_{s,u}} \label{eq:goal-maxsum-single-downlink} \\
\mbox{s.t. } 
& \sum_{g \in \mathcal{G}} x_t^{s,g} \le M_s, && \forall s \in \mathcal{S}, t=1,\ldots,T \label{eq:sat-trans-sch} \\
& \sum_{s \in \mathcal{S} } x_t^{s,g} \le R_g, && \forall g \in \mathcal{G},  t=1,\ldots,T \label{eq:ground-rec-sch} \\
%& \sum_{s \in \mathcal{S}} y_{s,u}   \ge \Lambda, && \forall u \in \mathcal{U}, \label{eq:demand-sch} \\
&  \sum_{t=1}^{T} n_t^{s,g}  x_t^{s,g} \ge \sum_{\forall u \text{ s.t. } g \in u} y_{s,u}, && \forall s\in \mathcal{S}, \forall g \in \mathcal{G} \label{eq:key-gen-sch}  \\
& x_t^{s,g} \in \{0,1\}, && \forall s \in \mathcal{S}, \forall g \in \mathcal{G},  t=1,\ldots,T \label{eq:x-cons} \\
& y_{s,u} \in Z, && \forall s \in \mathcal{S}, \forall u \in \mathcal{U} \label{eq:num-key-cons}
\end{align}

%After solving for $\Lambda$ as in the above, we can use the iterative method in \S\ref{sec:phase2-pairwise-keys} to use the leftover keys for the ground station pairs that are not yet saturated. 

In the above, (\ref{eq:goal-maxmin-single-downlink}) and (\ref{eq:goal-maxsum-single-downlink}) are the objective functions for Max-Min and Max-Sum optimization, respectively. Eq. (\ref{eq:sat-trans-sch}) represents the constraint on the number of transmitters for each satellite, while (\ref{eq:ground-rec-sch}) represents the constraint on the number of receivers for each ground station. Eq. (\ref{eq:key-gen-sch}) indicates that the total number of key bits that satellite $s$ generates for ground station $g$ needs to be no less than the number of key bits that is used to generate pairwise keys for all the pairs that include $g$. Last,  (\ref{eq:x-cons}) and (\ref{eq:num-key-cons}) specify that $x_t^{s,g}$ and $y_{s,u}$ are binary and integer variables, respectively. 

The above optimization problems are mixed-integer programming (MIP) problems, which can be solved using standard MIP solvers (e.g., CPLEX~\cite{CPLEX}). The total number of decision variables is $T |\mathcal{G}| |\mathcal{S}| + |\mathcal{S}| |\mathcal{U}|$. Even though we can remove some of the decision variables, e.g., $x_t^{s,g}$ if satellite $s$ is not in view of ground station $g$ in slot $t$, the number of decision variables can still be very large for a large number of satellites and ground stations. In \S\ref{sec:os-framework}, we propose a more efficient opportunistic scheduling framework.

%% file: framework.tex
\section{Opportunistic Scheduling Framework} \label{sec:os-framework}

\begin{figure}[t]%left, bottom, right, top
\centerline{\includegraphics[width=0.75\textwidth, trim = 0.0cm 7.0cm 0.0cm 0.0cm, clip]{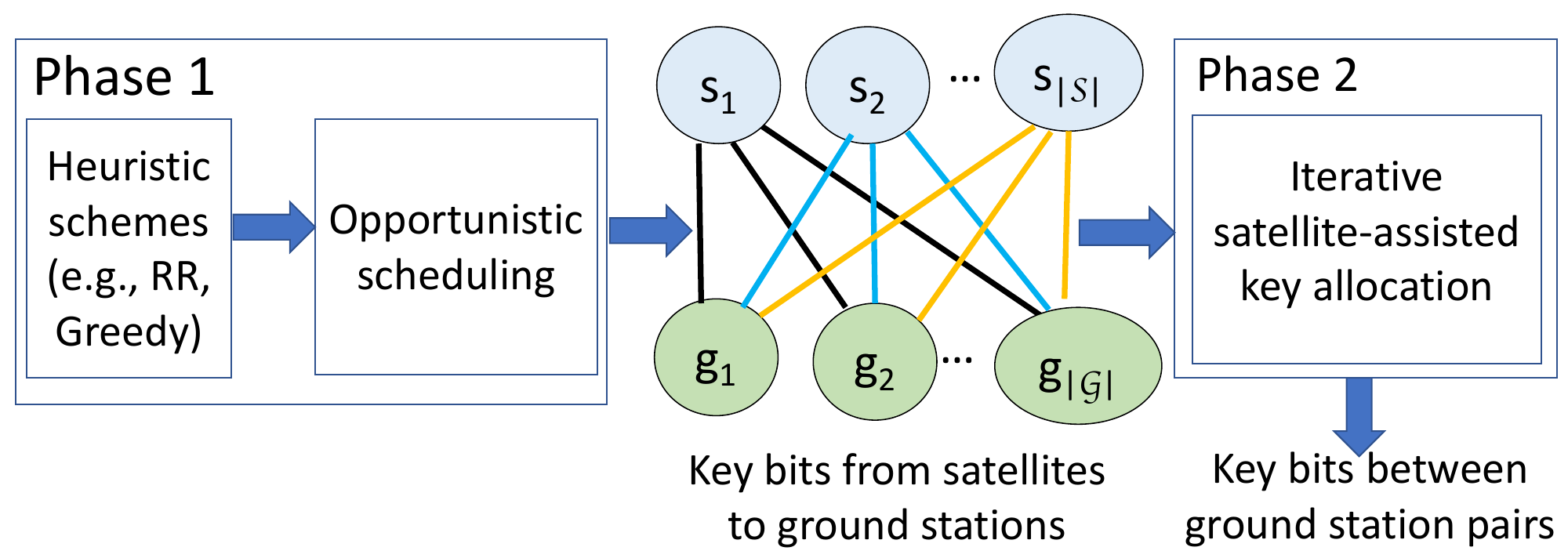}}
   %\vspace{-0.15in}
\caption{{Opportunistic scheduling framework. Phase 1 schedules satellites to establish keys with  individual ground stations, leading to key bits, $K_{s,g}$ between satellite $s$ and ground station $g$, $\forall s$, $\forall g$, colored in the figure based on the satellites. Phase 2 uses iterative optimization to establish keys among the ground station pairs.}}
\label{fig:framework}
%\vspace{-0.10in}
\end{figure}

Our proposed  opportunistic scheduling framework divides the problem 
%of key establishment for ground 
into two phases;
%, as illustrate in 
see Fig.~\ref{fig:framework}. Phase 1 assigns satellite to ground stations to determine $K_{s,g}$, $\forall s  \in \mathcal{S}$ and $g \in \mathcal{G}$ at the end of time $T$, and  Phase 2 further determines pairwise keys for each ground station pair through each satellite, i.e., $y_{s,u}$ given $K_{s,g}$, $\forall s  \in \mathcal{S}$, $g \in \mathcal{G}$, and $u \in \mathcal{U}$. These two phases are closely related: to achieve efficient and fair key establishment in Phase 2, the satellite assignment in Phase 1 needs to be efficient and fair. Consider an extreme case. Suppose that satellite $s$ generates a large number of key bits with only one ground station. Then all these key bits are useless in Phase 2, since satellite $s$ also needs to have key bits with other ground stations to be able to perform XOR operations to establish keys between $g$ and other ground stations. Therefore, in Phase 1, a natural goal is to achieve fair scheduling so that each satellite $s$ generate similar numbers of key bits across the ground stations.
%achieve fair assignments of the satellites across the ground stations. 

\BULLET {\bf Phase 1:} Opportunistic scheduling for QKD between satellites and ground stations. 
%However, satellite $s$ may not be able to serve some ground stations in the entire time interval. In addition, for the ground stations that $s$ can serve, the number of key bits that $s$ generates with the ground stations can vary significantly due to multiple factors such as the locations and dynamics in transmissivity and weather conditions. 
%
Consider a satellite, $s$. The number of key bits that $s$ generates with the ground stations can vary significantly due to multiple factors such as the locations and dynamics in transmissivity and weather conditions. Scheduling the satellite to serve the ground stations over time is akin to scheduling a base station to serve multiple users in wireless communication (e.g., cellular network systems), which has been extensively studied. We leverage {\em opportunistic scheduling}, a well-established framework in the area of wireless communication
(see survey~\cite{Asadi2013:opp-survey} and the references within), to solve this problem.  
The main idea of opportunistic scheduling is to dynamically assign transmission resources, e.g., time slots, subcarriers, or power, to users based on their instantaneous channel conditions. It exploits the natural variability in wireless channels across multiple users (e.g., due to fading, multipath, and mobility) to opportunistically serve users 
%, prioritizing users 
with favorable channel conditions
%the best current channel state 
to maximize overall system performance,
%(and hence the name ``opportunistic''), 
while potentially balancing trade-offs in QoS (e.g., delay, jitter) and fairness. 

The satellite scheduling problem in our setting, however, differs from the typical setup in that we need to schedule multiple satellites to serve a set of ground stations with constraints on the number of transmitters for the satellites and number of receivers for the ground stations. In addition, for one satellite, it is only in view of a subset of ground stations, or no ground station at all, for a large number of slots. We describe our  opportunistic satellite scheduling algorithm in \S\ref{sec:sat-scheduling}.

\BULLET {\bf Phase 2:} Satellite-assisted key assignments to ground station pairs. The second phase can be easily solved through a maxmin formulation 
%as follows.
%For fairness across ground station pairs, we present a maxmin formulation for key allocation. 
%The decision variables are 
with the decision variables $y_{s,u}$, $\forall s \in \mathcal{S}$, $u \in \mathcal{U}$, taking  $K_{s,g}$, $\forall s \in \mathcal{S}$, $g \in \mathcal{G}$ obtained by Phase 1 as the input.
\begin{align}
\text{maximize:} & \quad \min_{u} \sum_{s}  y_{s,u} \label{eq:goal-opt-phase2} \\
\mbox{s.t. } 
& \sum_{\forall u \text{ s.t. } g \in u} y_{s,u}  \le |K_{s,g}|, && \forall s \in \mathcal{S}, \forall g \in \mathcal{G}  \\
%
%wb, moved to objective function
%& \sum_{s}  y_{s,u} \ge \Lambda, && \forall u \in \mathcal{U} \\
& y_{s,u} \in Z, && \forall s \in \mathcal{S}, \forall u \in \mathcal{U} 
\end{align}
The above problem is also an MIP problem and can be solved using standard MIP solvers. It has $|\mathcal{S}||\mathcal{U}|$ decision variables and hence is much easier to solve than the formulation in (\ref{eq:goal-maxmin-single-downlink}). We solve the above problem iteratively. That is, after the first iteration, we remove the key bits that have been allocated, and the ground station pairs that have key bits equal to the obtained objective function value (i.e., no more keys can be established for them) so that they  are not considered in the next round. The process continues until no more keys can be established between any ground station pair.  
%Note that this is referring to the node pairs, not individual nodes. In other words, if $\sum_s y_{s,u}$ equals $\Lambda$ in round $i$, we can still consider $g \in u$ with other ground stations, it is just that we do not need to consider pair $u$ any more.

%%wb, 10/11/2025, old version
%What remains to be solved is scheduling satellites to serve ground stations in the first phase. As mentioned earlier, a natural goal in this phase is to achieve fair scheduling so that each satellite $s$ generate similar numbers of keys across the ground stations. 

%% file: os.tex
%\section{Opportunistic Satellite Scheduling} 
%\section{Opportunistic Satellite Scheduling to Ground Stations (Phase 1)}
%\section{Opportunistic  Scheduling for QKD between Satellites and Ground Stations}
\section{Phase 1 Opportunistic Scheduling}
\label{sec:sat-scheduling}

In this section, we detail our design of opportunistic satellite scheduling, i.e., the problem in Phase 1 in our proposed framework (see \S\ref{sec:os-framework}). For ease of exposition, we first consider the case of a single satellite serving a set of ground stations, and then extend the solution to a constellation of multiple satellites. Unless otherwise specified, we consider the realistic scenario where each ground station has a single receiver, i.e., $R_g=1$, and each satellite has a single transmitter, i.e., $M_s=1$; our approach can be easily extended to the more general case of multiple receivers and transmitters.

\subsection{Single Satellite}
 We model the single-satellite setting similar to the cellular network setting with one base station and multiple users, and time varying channel conditions in~\cite{Liu2001:oppor,Liu2003:oppor}. The main difference from the cellular network setting is that a ground station can only be served by the satellite for a small number of slots in a day.
%,  (e.g., 500 to 600 slots in 24$\times$3600 slots). 
Consider all the time slots $\{1,\ldots,T\}$. 
We can ignore the slots in which no ground stations can be served the satellite. 
%For the slots in which only a single ground station can be served by the satellite, clearly, the best option is for the satellite to serve that ground station. For the remaining slots, we need to design satellite scheduling algorithms.
Let $\mathcal{T}$ denote  the set of remaining slots, i.e., slots in which at least one ground station can be served by the satellite. The scheduling algorithm below only considers slots in $\mathcal{T}$.  

%wb, 10/4/2025, put t to be superscript, consistent with Liu2003 paper. Also, for the update rule, it is much easier if time is as superscript. Specifically, U_t^g -> U_g^t

To achieve high aggregate key rate, while maintaining certain degree of fairness among the ground stations, we use the {\em minimum-performance guarantee} scheduling framework in \cite{Liu2003:oppor}, which provides an absolute rate guarantee for each ground station. Specifically, let $r_g \ge 0$ denote the minimum average key rate for ground station $g$ over $\mathcal{T}$.
Let $U_g^t \ge 0$ denote the level of performance (or utility) obtained when the satellite serves ground station $g$ in slot $t$. For example, $U_g^t$ can be the number of key bits established between the satellite and the ground in slot $t$. 
Let 
$\vec{U}^t=\left(U^t_g, \forall g \in \mathcal{G}\right)$ 
represent the vector of key rate for all the ground station in slot $t$. 
%Assume that 
%wb, 10/14/2025
%When $\{\vec{U}^t\}$ is stationary and ergodic (which we will justify later),
For ease of exposition, we omit the superscript $t$ and let $\vec{U}=\left(U_g, \forall g \in \mathcal{G}\right)$ represent a vector of random variables, where $U_g$ represents the performance value of ground station $g$ in a generic time slot. A policy $Q$ determines which ground station is served by the satellite. 
%The minimum-performance guarantee scheduling framework finds 
%An optimal policy $Q^*$ that specifies which ground station that the satellite serves in each slot in $\mathcal{T}$ so that 

The problem of optimal minimum-performance guarantee scheduling is to find a policy 
%\begin{equation}
%\begin{split}
\begin{align}
\text{maximize}_{Q \in \Theta} & \quad \mathbb{E}\left( \sum_g U_g \mathbf{1}_{Q(\vec{U})=g} \right)= \sum_g  \mathbb{E}\left(U_g \mathbf{1}_{Q(\vec{U})=g} \right) \\
\text{subject to} & \quad \mathbb{E}\left(U_g \mathbf{1}_{Q(\vec{U})=g}\right) \ge r_g
\end{align}
where $\Theta$ represents the set of all stationary policies,
$\mathbf{1}_{Q(\vec{U})=g}$ is an indicator function, i.e., it is 1 if the policy selects ground station $g$, and 0 otherwise.
In other words, an optimal policy $Q^*$ satisfies that for each ground station $g$, its average key rate with the satellite is at least $r_g$ and the the total expected system performance is maximized.

%%wb, 10/4/2025, more detailed definition of the optimization problem; not sure we need that much details
\iffalse 
Let $
E\left(U_{Q(\vec{U})}\right)= E\left( \sum_g U_g \mathbf{1}_{Q(\vec{U})} \right)= \sum_g  E\left(U_g \mathbf{1}_{Q(\vec{U})} \right).  
$ 
That is, $E\left(U_{Q(\vec{U})}\right)$ represents the average system
performance value associated with policy $Q$.

The problem of optimal minimum-performance guarantee scheduling is to find a policy 
%%\mathbb{E}
\begin{equation}
\begin{split}
\text{maximize}_{Q \in \Theta} & \quad E\left(U_{Q(\vec{U})}\right) \\
\text{subject to} & \quad E\left(U_g \mathbf{1}_{Q(\vec{U})=g}\right) \ge r_g
\end{split}
\end{equation}
where $\Theta$ represents the set of all stationary policies,
%feasible policies (i.e., satisfying the minimum key rates for all the ground stations),  
$\mathbf{1}_{Q^*(\vec{U})=g}$ is an indicator function, i.e., it is 1 if the policy selects ground station $g$, and 0 otherwise.
\fi 

%
The study in~\cite{Liu2003:oppor} identifies an optimal policy as 
\begin{equation}
Q^*(\vec{U}) = \arg\max_g [(1+\lambda_g^*) U_g]   \label{eq:Q-single} 
\end{equation}
where the $\lambda_g^*$'s are real parameters  satisfying (i) $\min_g (\lambda_g^*)=0$; (ii) For all $g \in \mathcal{G}$, $\mathbb{E}\left(U_g \mathbf{1}_{Q^*(\vec{U})=g}\right) \ge r_g$; and (iii) For all $g \in \mathcal{G}$, if $\mathbb{E}\left(U_g \mathbf{1}_{Q^*(\vec{U})=g}\right) > r_g$, then $\lambda_g^*=0$, .  

Intuitively, the above policy selects the relatively-best ground station, i.e.,  ground station $g$ that has the highest $(1+\lambda_g^*) U_g$, where the non-negative parameters $\lambda_g^*$'s  scale the performance values. This policy increases the chance that the ground stations under unfavorable conditions are selected, so that their minimum rate guarantee is achieved, i.e., satisfying $\mathbb{E}\left(U_g \mathbf{1}_{Q^*(\vec {U})=g}\right) = r_g$.
%for those ground stations.
%their average rates equal to their respective minimum rates); 
For the favorable ground stations, i.e., those with $\mathbb{E}\left(U_g \mathbf{1}_{Q^*(\vec {U})=g}\right) > r_g$, they get higher rates than their pre-determined minimum rates, taking advantage their higher key rates.  

The optimal policy in (\ref{eq:Q-single}) requires estimating parameters, $\lambda_g^*$, for each ground station $g$. Let vector $\vec{\lambda^*} = (\lambda_g^*, \forall g\in \mathcal{G})$ represent the scalars for all the ground stations. 
Based on the conditions that
%for $\lambda_g^*$ that 
%described earlier, specifically, 
 if $\lambda_g^*>0$, then $\mathbb{E}\left(U_g \mathbf{1}_{Q^*(\vec{U})=g}\right) = r_g$,  we see that $\vec{\lambda}^*$ is a root of $f(\vec{\lambda})=0$; the component for satellite $g$ in $f(\vec{\lambda})$ is defined as   
\[
f_g(\vec{\lambda})=(\lambda_g - \min_{g'}(\lambda_{g'}))\left(\mathbb{E}\left(U_g \mathbf{1}_{Q^*(\vec{U})=g}\right) - r_g\right)\,. 
\]

Following the approach in~\cite{Liu2003:oppor}, we use stochastic approximation to obtain the root $\lambda_g^*$ through a sequence of iterates, $\lambda_g^1,\lambda_g^2, \ldots$. Each iterate $\lambda_g^t$ defines a policy for slot $t$ as 
\begin{equation}
Q^t(\vec{U}^t)=\arg\max_g [(1+\lambda_g^t) U_g^t]. \label{eq:Q-single-t}
\end{equation}
The update rule is
    \begin{equation}
        \lambda_g^{t+1} = \left[ \lambda_g^t - \delta^t \left( U_g^t \mathbf{1}_{Q^t(\vec{U}^t)=g} - r_g  \right) \right]^+ 
        \label{eq:single-update}
    \end{equation}
where $\delta^t$ is a small positive step size that controls the update rate;  
$U_g^t \mathbf{1}_{Q^t(\vec{U}^t)=g}$ is $U_g^t$
%, the key rate of ground station $g$ in slot $t$ 
if policy $Q^t$ schedules the satellite to serve $g$ in slot $t$, and 0 otherwise; $r_g$ is the pre-determined minimum performance, and 
$[x]^+ = \max(x, 0)$ ensures that \( \lambda_g^{t+1} \geq 0 \).

{\bf Practical considerations.} We set the initial value, $\lambda_g^{1}=0$, $\forall g \in \mathcal{G}$. For simplicity, the update step size, $\delta^t$, is set to a small constant, 0.01. We further normalize all the key rates, $U_g^t$, and the predetermined minimum key rate, $r_g$, %$\forall g \in \mathcal{G}$, 
by the maximum key rate of all the ground stations served by the satellite over all the slots, so that they are in $[0,1]$. The update rule in (\ref{eq:single-update}) requires a long sequence of time slots to reach convergence. We repeat time slots in $\mathcal{T}$  multiple times %(simulating multiple days 
until convergence.
%$\lambda_g^{t}$'s converge. 
We then use the policy for the last intervals of $\mathcal{T}$ as the optimal opportunistic policy. 
%Since we simply repeat $\mathcal{T}$, the key rate vector $\{\vec{U}^t\}$ is stationary and ergodic.

%In a slot $t$ where only a single ground station can be served by the satellite, based on the policy in Eq. (\ref{eq:Q-single-t}), the satellite will serve this ground station, as expected. To ensure convergence, we repeat time slots in $\mathcal{T}$ over $n$ days until $\lambda_g^{t}$ converges (determined as \bing{xx}). We then use the policy for $n$ as the optimal opportunistic policy. Since we simply repeat $\mathcal{T}$, the key rate vector $\{\vec{U}^t\}$ is stationary and ergodic.

\subsection{Multiple Satellites} \label{sec:op-multi-scheduling}

We now consider the more general case of multiple satellites serving a set of ground stations. Let $\mathcal{T}_s$ denote the set of slots for satellite $s$ in which at least one ground station can be served by satellite $s$. Let $\mathcal{T}=\bigcup_{s \in \mathcal{S}} \mathcal{T}_s$. We  consider the slots in $\mathcal{T}$ for multi-satellite scheduling. 
Let $r_{s,g}$ denote the pre-determined minimum key rate for ground station $g$ served by satellite $s$.
Let $U_{s,g}^t \ge 0$ denote the key rate when satellite $s$ serves ground station $g$ in slot $t$. 
Let 
$\vec{U}_s^t=\left(U^t_{s,g}, \forall g \in \mathcal{G}\right)$ 
represent the vector of key rates for all the ground stations when served by satellite $s$ in slot $t$. 

Simply treating all the satellites independently, we define policy $Q_s^t$ for satellite $s$ and scalar $\lambda_{s,g}^t$ as in the single-satellite case. Then directly following (\ref{eq:Q-single-t}) and (\ref{eq:single-update}), the policy for each satellite $s$ in $\forall t \in \mathcal{T}_s$ is 
\begin{equation}
Q_s^t(\vec{U}_s^t)=\arg\max_g [(1+\lambda_{s,g}^t) U_{s,g}^t]. \label{eq:Q-multi-t}
\end{equation}
The update rule is
    \begin{equation}
        \lambda_{s,g}^{t+1} = \left[ \lambda_{s,g}^t - \delta^t \left( U_{s,g}^t \mathbf{1}_{Q_s^t(\vec{U}_s^t)=g} - r_{s,g}  \right) \right]^+\,. 
        \label{eq:multi-update}
    \end{equation}
The above policy, however, does not account for the coupling of the satellites in multi-satellite scenarios. 
%nstraint that each ground station has a single receiver, and hence can only be served by a single satellite. 

\input{multi-sate-fig}

Two examples are shown in Fig.~\ref{fig:bipartite}. Let $w_{s,g}^t \coloneq (1+\lambda_{s,g}^t) U_{s,g}^t$. We refer to $w_{s,g}^t$ as the {\em weight} for satellite $s$ and ground station $g$ in slot $t$, represented as a line connecting $s$ and $g$; the thicker the line, the larger weight. 
%which are represented by the thickness of the lines in 
Figures~\ref{fig:bipartite}a and b show the actions of two satellites, $s_1$ and $s_2$, are coupled since both of them can serve ground station $g_2$. 
In Fig.~\ref{fig:bipartite}a, following (\ref{eq:Q-multi-t}), satellites $s_1$ and $s_2$ serve ground stations $g_1$ and $g_2$, respectively. In Fig.~\ref{fig:bipartite}b, however, the weight between $s_1$ and $g_2$ is the largest. If $s_1$ serves $g_2$, then $g_1$ will not be served in that slot. An alternative strategy schedules $s_1$ to serve $g_1$, and $s_2$ to serve $g_2$. It leads to a more balanced key rate among the ground stations, and hence more desirable for generating pairwise keys among ground station pairs in Phase 2 (see \S\ref{sec:os-framework}), even though the total key rate is lower than the of the first strategy.   

\subsubsection{Multi-satellite Scheduling}
%over Bipartite Graph} 
\label{sec:multi-sate-bipartite}

In general, we model the multi-satellite scheduling problem as a bipartite graph. In each slot, we consider the set of satellites that can serve at least one ground station, and the set of ground stations that can be served by at least one satellite. The connection between satellite $s$ and $g$ is marked as the weight, $w_{s,g}^t$. Scheduling the satellites can then be modeled as a bipartite matching problem. 

Let $N_S$ and $N_G$ denote respectively the number of satellites and ground stations in the graph (we omit the superscript $t$ for clarity). We may have $N_S<N_G$, $N_S=N_G$, or $N_S>N_G$. %Instead of Maximum weight matching, which can lead 
Since $N_S$ may not equal $N_G$, some satellites may not be scheduled to serve any ground station if $N_S > N_G$, or a ground station may not be served if $N_S < N_G$. 
As illustrate earlier, since more balanced keys among the ground stations and satellites can lead to more balanced pairwise keys across the ground stations, we want to use all the satellites if $N_S \le N_G$, or serve all the ground stations if $N_S \ge N_G$. 
%Instead of Maximum weight matching, 
We model the problem as a two-dimensional (2D) assignment problem, also
known as the linear sum assignment problem~\cite{Burkard2009:assignment}. In the following, we describe the problem assuming $N_G \le N_S$; the other case can be simply modeled by switching satellite and ground stations. 
%wb, 10/11/2025, swtich to $N_G \le N_S$, which is the common case in simulation

Consider a $N_G \times N_S$ matrix $\mathbf {W}$ of weights, where the weights are as defined above. The 2D
rectangular assignment problem consists of choosing one
element in each row and at most one element in each
column (i.e., using one satellite to serve a ground station) such that the sum of the chosen elements maximized. Specifically, it solves an optimization problem
\begin{align}
& \mathbf{X}^* = \arg \max_{\mathbf{x}} \sum_{i=1}^{N_G} \sum_{j=1}^{N_S} w_{i,j} x_{i,j}  \\
\text{subject to} & \quad \sum_{j=1}^{N_S} x_{i,j} = 1,  && \forall i=1,\ldots, N_G \\
& \quad  \sum_{i=1}^{N_G} x_{i,j} \le 1,  && \forall j= 1,\ldots, N_S\\  
& x_{i,j} \in \{0,1\}, && \forall i=1,\ldots, N_G,  \forall j= 1,\ldots, N_S 
\end{align}
The optimal solution $\mathbf{X}^*$ determines the schedule for the slot.
%scheduling of the satellites to the ground stations in the slot. 

The above 2D assignment problem can be solved efficiently using polynomial algorithms. When $N_S=N_G=N$, it can be solved using the Hungarian algorithm with the complexity of $O(N^3)$~\cite{Burkard2009:assignment}. For the more general case, it can be solved using the JVC algorithm~\cite{Jonker1987:JV, Drummond1990:2D}. We use \texttt{linear\_sum\_assignment} in  \texttt{scipy.optimize} to solve it, which uses a modified JVC algorithm~\cite{Crouse2016:2D}.

After obtaining the schedule for satellite $s$ in slot $t$, we use the update rule in (\ref{eq:multi-update}) to  obtain $\lambda_{s,g}^{t+1}$ for the next slot. We again set the initial value, $\lambda_{s,g}^{1}=0$, $\forall s, \forall g$, and set $\delta^t$ to 0.01.
%The small constant $\delta^t$ is set to 0.01. 
The key rate values $U_{s,g}^t$ and $r_{s,g}$ are normalized by the maximum key rate of all the ground stations and satellites 
over all the slots.
%so that they are in [0, 1]
We again repeat time slots in $\mathcal{T}$ multiple times until convergence, and take the policy in the last $\mathcal{T}$ as the schedule.

%\bing{Comment about minimum rate guarantee: show empirically; matching makes it difficult, analytical results left as future work.}
In summary, the above multi-satellite scheduling algorithm combines single-satellite scheduling and 2D assignment to determine the schedule for each satellite in each slot. Because of the 2D assignment, the schedule for a satellite does not necessarily follow (\ref{eq:Q-multi-t}) and it is not clear whether the minimum rate $r_{s,g}$ is guaranteed for satellite $s$ and ground station $g$. In \S\ref{sec:eval}, we show, empirically, that the minimum rate is satisfied for the above algorithm
%after running multi-satellite opportunistic scheduling 
for all the cases that we evaluate. Further analytical study is left as future work.

%Our goal is to guarantee that the minimum key rate $r_{s,g}$ is satisfied while maximize the sum of the key rate. 

%As in Eq. (\ref{eq:Q-single-t}), we define $\lambda_{s,g}^t$ to scale the key rate for satellite $s$ with ground station $g$. However, we cannot schedule the satellites independently as in Eq. (\ref{eq:Q-single-t}), since two satellites can choose the same ground station to serve, which violates the constraint that each ground station has a single receiver and hence can only be served by a single satellite.

%%wb, 10/5/2025, motivate minimum-cost perfect matching in bipartite graph, 
%We propose a minimum-cost matching formulation to resolve the above problem. Specifically, for slot $t$, we consider a bipartite graph as 

%\input{sigmetrics/os-single}
%\input{sigmetrics/os-multi}

\input{min-rate}

%% file: multi-sate-fig.tex
\begin{figure}[t]%left, bottom, right, top
   \centerline{\includegraphics[width=0.75\textwidth, trim = 0.0cm 7.3cm 0.0cm 1.0cm, clip]{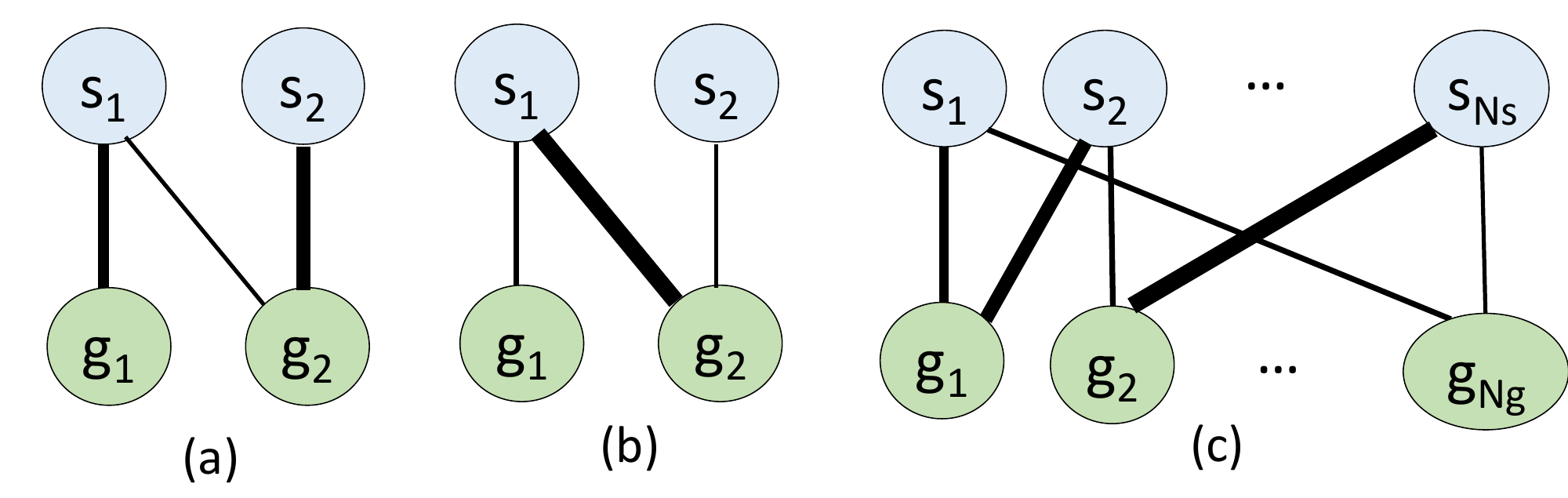}}
   %\vspace{-0.15in}
\caption{{Bipartite graph representing the relationship between satellites and ground stations in one time slot. A satellite is connected to a ground station if it can serve the ground station in that slot. A thicker line connecting satellite $s$ and ground station $g$ represents a higher weight, $w_{s,g}^t$. In (a) and (b), we show two examples that illustrate the coupling of the satellites; a general case is shown in (c).}}
\label{fig:bipartite}
%\vspace{-0.10in}
\end{figure}

%% file: min-rate.tex
\subsubsection{Predetermining Minimum Rates} \label{sec:min-rate-heuristics}
%\subsection{Round-robin Heuristic}

We next design two heuristic algorithms, Round-Robin (RR) and Greedy heuristic, to determine the key rate, $r_{s,g}$, $\forall s, \forall g$, beforehand, which is used as input to our opportunistic scheduling algorithm.
%specifically, the update rule in (\ref{eq:multi-update}). 

\BULLET {\bf RR.}  
%which is often used to obtain fair resource allocation to multiple parties. 
%Let $\mathcal{T}_j$ be the set of slots in which at least one ground station can be served by the satellite $s_j$. 
Recall that $\mathcal{T}$ represents the set of slots in which at least one satellite can serve at least one ground station.  We design RR so that each satellite serves each ground station for a similar number of slots. Specifically, we keep a counter $C_{s,g}$, which denotes the number of slots for which satellite $s$ has served ground station $g$ at the end of $\mathcal{T}$, $\forall s$, $\forall g$. All the counters are initialized to zero.  
We then find all the slots in $\mathcal{T}$, in which a single ground station, $g$, can only be served by a single satellite, $s$, and $s$ is not in view of any other ground stations. For these slots, we directly schedule
satellite $s$ to serve $g$,  and increment the counter, $C_{s,g}$, accordingly. After that, we consider the
remaining slots in $\mathcal{T}$ sequentially. For each of these slots, we consider a weighted bipartite graph, where a satellite $s$ is connected to a ground station $g$ if $s$ can serve $g$ in that slot, and the weight of the edge is set to the current value of  $C_{s,g}$.
%, which is the current count that ground station $g$ has been served by satellite $s$ up to that slot. 
To prioritize the satellite and ground station pairs that have lower count, we use 2D assignment (as in \S\ref{sec:multi-sate-bipartite}), with the goal of minimizing the sum of the weights. The output of the  2D assignment determines which satellite serves which ground station in each slot. 

\BULLET {\bf Greedy heuristic.} We design this heuristic to prioritize key rate, with some consideration of fairness. Specifically, in each slot $t$, let $S^t_g$ denote the set of satellites that can serve ground station $g$. This greedy heuristic chooses the satellite $s \in S^t_g$ that has the highest key rate with $g$ to serve $g$. A {\em conflict} occurs if the same satellite is the best satellite for two ground stations, $g$ and $g'$, since a satellite can only serve a single ground station at one point of time (assuming each satellite has a single transmitter). In this case, the satellite $s$ is assigned to the ground station with which $s$ has fewer key bits so as to balance the number of key bits between $s$ and the ground stations. 
This heuristic may cause a satellite $s$ to generate many key bits with one ground station, but only a small number of key bits with other ground stations, which will limit the number of key bits that can be generated for ground station pairs through $s$ in Phase 2.

%Since this heuristic is greedy in nature, it can cause a satellite $s$ to generate many key bits with a ground station, $g$, while only generate a small number of key bits with another ground station, $g'$. As a result, ground stations $g$ and $g'$ are not able to generate many key bits with each other through $s$ in Phase 2,  limiting the performance of this scheme, as we shall see in \S\ref{sec:eval}.  

%Compared to RR, this Greedy heuristic has less considerations of Phase 2 since a satellite $s$ may be preferred by a ground station, $g$, but not by another ground station, $g'$, which limits the size of the key that $g$ and $g'$ can generate through $s$ in Phase 2. 

%has high key rate with a ground station, $g$, may have very low key rate with another ground station, $g'$, which  

%(i) max-weight matching, where we find a matching between the satellites and ground stations with the weight of each line marked as $-C_{i,j}^t$, where $C_{i,j}^t$ is the current count that ground station $g_i$ has been served by satellite $s_j$ up to slot $t$. (ii) greedy heuristic, where we simply pick the smallest $C_{i,j}^t$, remove ground station $g_i$ and satellite $s_j$ from the graph until no ground station can be served.   

For both of these heuristics, once the satellite schedules are determined, we can obtain the total number of key bits that satellite $s$ generates with ground station  $g$ over $\mathcal{T}_s$, i.e., the subset of time slots in $\mathcal{T}$ in which satellite $s$ is in view of at least one ground station. We then obtain the average key rate as the total number of key bits divided by $|\mathcal{T}_s|$, which is used as $r_{s,g}$ for multi-satellite opportunistic scheduling. 

%the above multi-satellite formulation.  

%% file: results.tex
\section{Performance Evaluation} \label{sec:eval}

\input{simu-setup}

\input{need-for-scheduling}

\input{no-cloud}

\input{with-cloud}

\input{small-set-res}

%% file: simu-setup.tex
\subsection{Evaluation Setup} \label{sec:sim-setup}
We consider a polar constellation of LEO satellites in 20 rings, each ring with 20 satellites for a total of 400 satellites as  in~\cite{Khatri21:Spooky,Panigrahy22:optimal}. Let $A$ denote the attitude of the satellites.
We set $A$ to 500 km, 800 km, or 1000 km. 
%The orbit time (the amount of time for a satellite to finish one orbit) is 5668, 6044, and 6298 seconds, for satellite altitudes of 500km, 800km and 1000km, respectively. 
Each satellite is equipped with a photon source that operates at a one GHz rate, i.e., generating $10^9$ photons per second.
We consider two QKD deployment scenarios: {\em global} deployment with 11 ground stations in 5 continents and {\em regional} deployment with 4 ground stations in North America. Specifically, the 11 ground stations in 5 continents include New York City (NYC), Washington D.C. (DC), Toronto, Houston, and Boston in North America, London and Dublin in Europe, Singapore and Mumbai in Asia, Sydney in Australia, and Johannesburg in Africa. They represent an envisioned global QKD system that provides inter-continental coverage. The 4 ground stations in North America include NYC, DC, Toronto, and Houston, which is a smaller-scale QKD system that is easier to deploy.
%and can provide coverage in a continent.
%
%a country or within a continent.
%may be deployed for economic and political benefits.   
%
%ground stations located in 11  cities in 5 continents: North America (New York City (NYC), Washington D.C. (DC), Toronto, Houston, Boston), Europe (London, Dublin), Asia (Singapore, Mumbai), Australia (Sydney), and Africa (Johannesburg). 
%Unless otherwise specified, our results are for this set of ground stations, involving 55 ground station pairs, representing a global deployment of a satellite-based QKD system. 
%To provide additional insights, we further show the results for four ground stations in North America (NYC, DC, Toronto, and Houston), which represents a regional deployment of QKD system, at the end. 
%
Considering the constraints of the current technologies, we focus on the setting where each satellite has a single transmitter and each ground station has a single receiver, i.e., $M_s=1$, $R_g=1$, $\forall s \in \mathcal{S}$ and $\forall g \in \mathcal{G}$. %and hence can only perform QKD with another ground station in each slot. 
%We also explore the case when each satellite has three transmitters, i.e., $M_s=3$; the results are very close to those when $M_s=1$ due to constraints of the ground stations. Henceforth, we only present the results when $M_s=1$.

We consider four days, the 15{th} day of March, June, September, and December, that represent different seasons of a year. Specifically, the channel models follow the actual weather and background photon measurement data for these four days in 2022, as described in  \S\ref{sec:model}. For each day, we obtain the satellite schedules for each slot of one second. Therefore, one day contains $T=86,400$ slots. 
The elevation angle threshold is set to $\theta=20$$^{\circ}$, i.e., a satellite 
%can only be in view of and 
can only serve a ground station when the elevation angle to the ground station is larger than 20$^{\circ}$.

%We compare our three proposed approaches with a baseline  scheduling strategy that only aims to maximize the total number of keys in each slot, with no fairness considerations. Specifically, it differs from slot-based weighted-sum in that the objective function is maximizing  $\sum_{s \in \mathcal{S}} \sum_{g,g' \in \mathcal{G}} x_t^{s,g,g'} n_t^{s,g,g'}$. Henceforth, we refer to this baseline strategy as {\em slot-based max-key}. The solutions of all the strategies are obtained using CPLEX~\cite{CPLEX}, a widely used MIP solver.

{\bf Performance metrics.} We evaluate the various scheduling strategies using two performance metrics: (i) {\em minimum key size}, i.e., the minimum number of secret key bits that are generated for the ground station pairs at the end of a day, and (ii) {\em total key size}, i.e., the sum of the secret key bits generated for all ground station pairs at the end of a day. The first metric represents fairness among the ground station pairs, while the second one represents overall system throughput. A desirable scheduling algorithm achieves an effective balance between these two metrics.
%leads to a good tradeoff of the above two metrics. 

{\bf Comparison schemes.} We evaluate two opportunistic scheduling schemes, {\em Op-RR} and {\em Op-Greedy}. They use respectively the outputs of the two heuristic schemes, {\em Round-Robin (RR)}, and {\em Greedy heuristic}, as input to opportunistic scheduling in Phase 1, and then the iterative optimization in (\ref{eq:goal-opt-phase2}) for Phase 2. We compare them with the results when using RR and Greedy for Phase 1 directly (i.e., without opportunistic scheduling) and then the same iterative optimization approach for Phase 2. In addition, we  compare them with the results from the {\em Max-Min} formulation, which is optimal in terms of the minimum key size metric, and the {\em Max-Sum} formulation, which is optimal in terms of the  total key size metric, see \S\ref{sec:opt}. Both Max-Min and Max-Sum rely on solving large-scale MIP problems. We obtain results for these problems using CPLEX~\cite{CPLEX} running on a high-performance computing facility with over 300 GB memory, and each problem instance takes multiple hours to complete. The two opportunistic scheduling schemes are much more scalable, taking only minutes to run, with little memory requirement.

%number of generated secret keys and minimum number of keys across all the ground station pairs at the end of a day.
%and (ii) minimum number of keys across the ground station pairs. 

%{\bf Comparison schemes.} We consider {\em Max-Min} and {\em Max-Sum} (see \S\ref{sec:opt}), which are optimal in terms of one of the above two metrics, two heuristic schemes, {\em Round-Robin (RR)}, and {\em Greedy heuristic}, which use heuristics for satellite scheduling in the first phase (see \S\ref{sec:min-rate-heuristics}) and then the formulation in (\ref{eq:goal-opt-phase2}) for the second phase, and two opportunistic scheduling schemes, {\em Op-RR} and {\em Op-Greedy}, that use RR and Greedy to obtain the feasible solution, respectively. 

%% file: need-for-scheduling.tex
\subsection{Need for Satellite Scheduling}
%\bing{Show 3 plots: histogram of the number of ground stations that a satellite can serve serve in a slot, $A=500$, 800, and 1000km.}

%\bing{Show 3 plots: for a ground station, histogram of the number satellites from which it can choose to be served in a slot, $A=500$, 800, and 1000km.}

\input{fig-need-for-sche}

%We first show the need for satellite scheduling. 
For both the global and regional QKD settings, a satellite can be in view of multiple ground stations, and a ground station can be in view of multiple satellites in a slot, hence presenting the need for scheduling satellites. 
%Weather conditions are ignored in the plots since whether 
Fig.~\ref{fig:need-for-sche} shows the distribution plots for the global QKD setting for the day in September; the results the other
three days are similar. 
The top row 
%of Fig.~\ref{fig:need-for-sche} 
shows the histogram of the number of ground stations that a satellite can choose to serve in a slot, considering all the satellites and slots.
%(the results are not sensitive to the season). 
The results for $A=500$, 800, and 1000 km are shown in the figure.  
%The y-axis represents the number of instances, considering all the $20\times20$  satellites in the constellation and all the slots in a day. 
We see a significant number of instances when a satellite can choose from 
multiple ground stations 
to serve in a slot, and the number of choices increases with altitude. Specifically, the maximum number of ground stations that a satellite can choose from is 4 when $A=500$ km, and 5 when $A=800$ and 1000 km.
%due to the relative locations of the ground stations. 
The bottom row of Fig.~\ref{fig:need-for-sche} shows the histogram of the number of satellites that can serve a ground station in a slot, considering all the ground stations and slots in a day. Only when $A=500$ km, a ground station is sometimes not in view of any satellite in one slot (and hence cannot be served in that slot); for $A=800$ km, a ground station can be served by 2 to 8 satellites, which is increased to 3 to 11 satellites when $A=1000$ km. The above results demonstrates the need for satellite scheduling.
%and the bipartite formulation in \S\ref{sec:op-multi-scheduling}. 
The regional QKD setting exhibits similar trends as in Fig.~\ref{fig:need-for-sche}, with slightly less ground stations (up to 4) that a  satellite can choose to serve, and less satellites (up to 9) that can serve a ground station in a slot. 
%\bing{It looks like that most of the time, the bipartite graph has more satellites than ground stations. Is that right?}

We next briefly describe the statistics regarding Phase 1 opportunistic scheduling. Recall that $\mathcal{T}_s$ represents the slots that satellite $s$ are considered in the scheduling. 
%wb, 10/13/2025, simplify to only describe the average values
For the global QKD setting, the average  $|\mathcal{T}_s|$ values are 3537, 6734, and 8999 when $A=500$, 800, and 1000 km,  respectively. For the regional QKD setting, the values are lower, as 1130, 1952, and 2487.
%\bing{For the global QKD setting, the average  $|\mathcal{T}_s|$ value is 3537, 6734, and 8999 when $A=500$, 800, and 1000 km,  respectively. For the regional QKD setting, $|\mathcal{T}_s|$ is lower, with the average  $|\mathcal{T}_s|$ value as 1130, 1952, and 2487 when $A=500$, 800, and 1000 km,  respectively.}
%
%{For the global QKD setting, $|\mathcal{T}_s|$ varies from 2125 to 4760 when $A=500$ km, for $A=800$ km and $1000$ km, the corresponding ranges are 5988 to 7555, and 8136 to 9959, respectively. For the regional QKD setting, $|\mathcal{T}_s|$ is lower, 
%since there are less ground stations, the values are lower, varying from 671 to 1612, 1563 to 2552, and 1712 to 2905 for $A=500$, 800, and 1000 km, respectively.} 
In each slot, opportunistic scheduling considers a bipartite graph connecting a subset of satellites and ground stations, with an edge connecting a satellite and ground station if that satellite can serve the ground station. For the global QKD setting, when $A=500$ km, the number of ground stations in a slot varies from 9 to 11  (i.e., most or all of the ground stations can be served by at least one satellite), and the number of satellites is larger, varying from 11 to 21; for $A=800$ and 1000 km, the number of ground stations is 11 per slot, and the number of satellites varied from 24 to 38, and 35 to 49, respectively. For the regional QKD setting, when $A=500$ km, the number of ground stations in a slot varies from 3 to 4 (i.e., most or all of the ground stations can be served by at least one satellite), and the number of satellites varies from 2 to 8; for $A=800$ and 1000 km, the number of ground stations is 4 per slot, and the number of satellites varies from 6 to 12, and 9 to 14, respectively.

%\bing{Say a bit about $|\mathcal{T}_s|$ for each satellite. Also, for the slots in $\mathcal{T}$, number of satellites and ground stations in each slot.}  

%Fig.~\ref{fig:Sat_Dist}b shows the corresponding results from the perspective of the ground stations, i.e., for a ground station, the number of satellites from which it can choose to be served, considering all the ground stations and slots in a day. We see a significant number of instances in which a ground station can be served by xxx satellites. Fig.~\ref{fig:Sat_Dist}c and d show the corresponding results for $A=1000$km, and again for the day in September. We see more choices for this higher altitude.
%: there are a large number of  instances in which a satellite can choose from 3 or 6 ground station pairs in a slot, while a ground station pair can choose from up to 8 satellites in a slot. 
%The results for other settings (figures omitted) also show similar t.
%Since a satellite has a single transmitter, it can choose one ground station pair to serve in a slot; since a ground station only has a single receiver, it can only be served by one satellite, and can only be in one pair that is served. Therefore, a satellite scheduling strategy is needed that considers the possible set of scheduling choices, and selects the one that satisfies a certain optimization goal. 

%% file: fig-need-for-sche.tex
\begin{figure*}[htbp]
    \centering
    % ---- Subfigure (a) ----
    \begin{subfigure}[b]{0.32\textwidth}
        \centering
        \includegraphics[width=\textwidth]{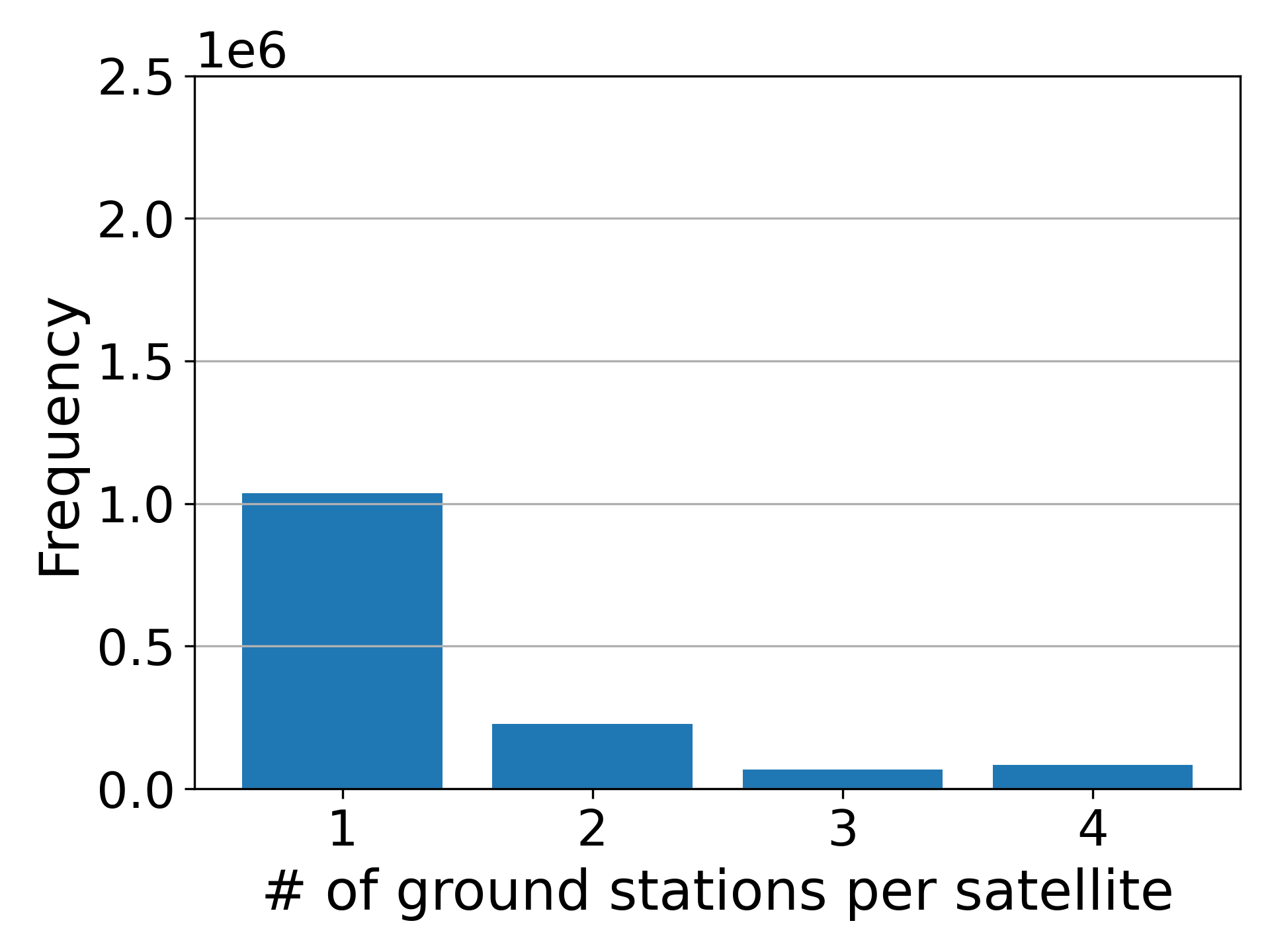}
        \caption{$A=500$ km, satellite view.}
        \label{fig:GS_Dist_500}
    \end{subfigure}
    % ---- Subfigure (b) ----
    \begin{subfigure}[b]{0.32\textwidth}
        \centering
        \includegraphics[width=\textwidth]{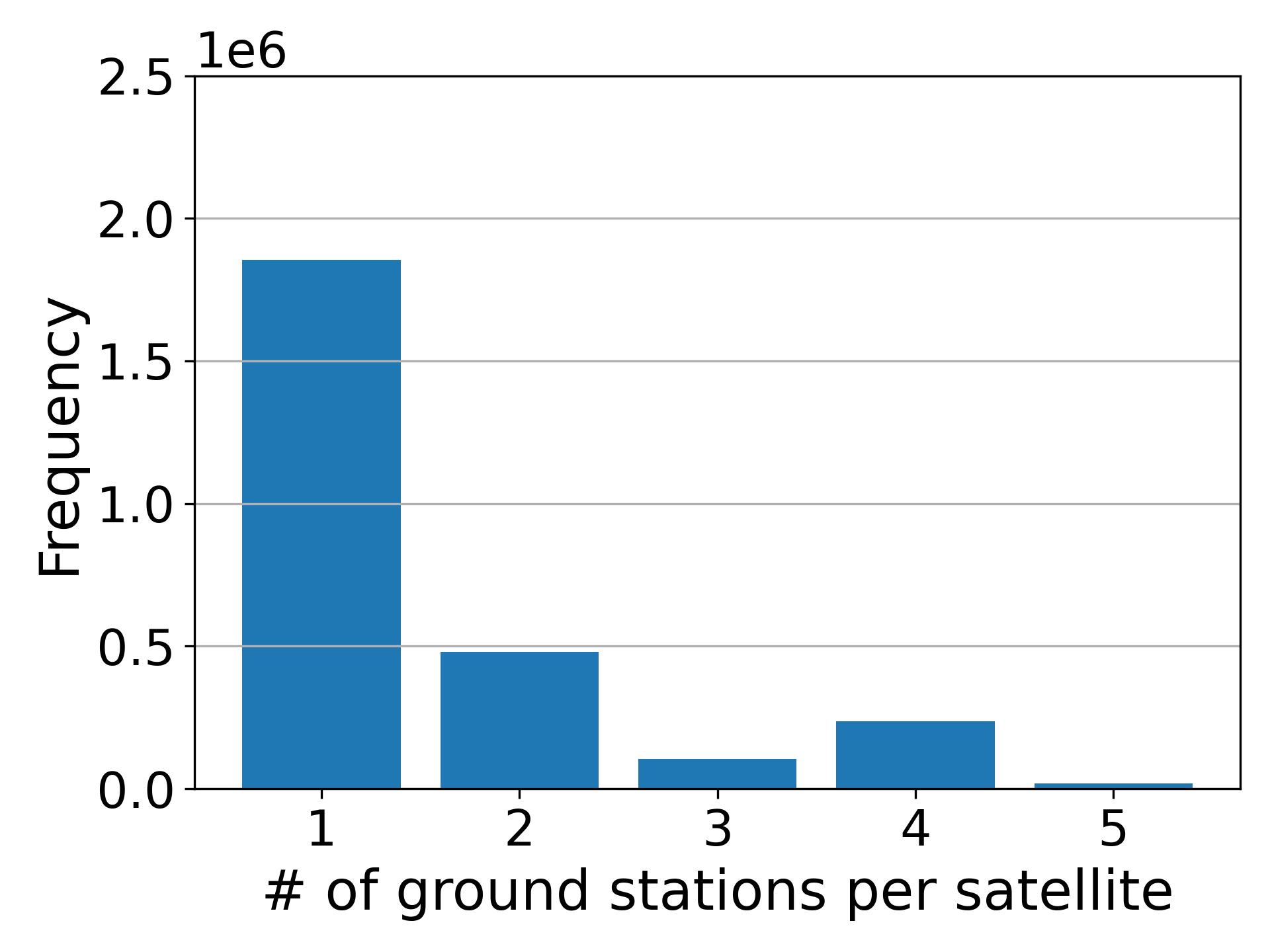}
        \caption{$A=800$ km, satellite view.}
        \label{fig:GS_Dist_800}
    \end{subfigure}
    % ---- Subfigure (c) ----
    \begin{subfigure}[b]{0.32\textwidth}
        \centering
        \includegraphics[width=\textwidth]{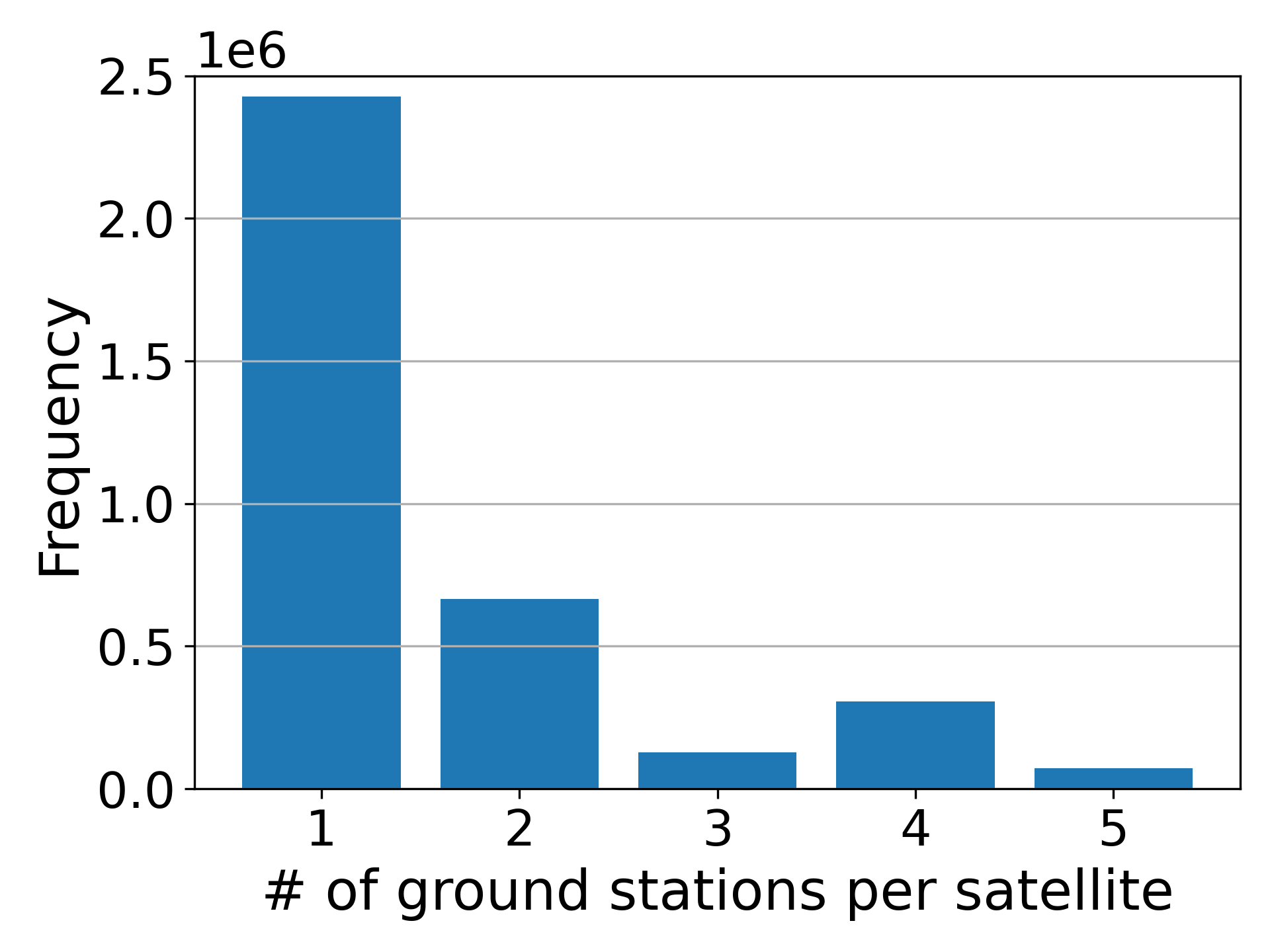}
        \caption{$A=1000$ km, satellite view. }
        \label{fig:GS_Dist_1000}
    \end{subfigure} \\
    %ground station view
    \vspace{0.10in}
    \begin{subfigure}[b]{0.32\textwidth}
        \centering
        \includegraphics[width=\textwidth]{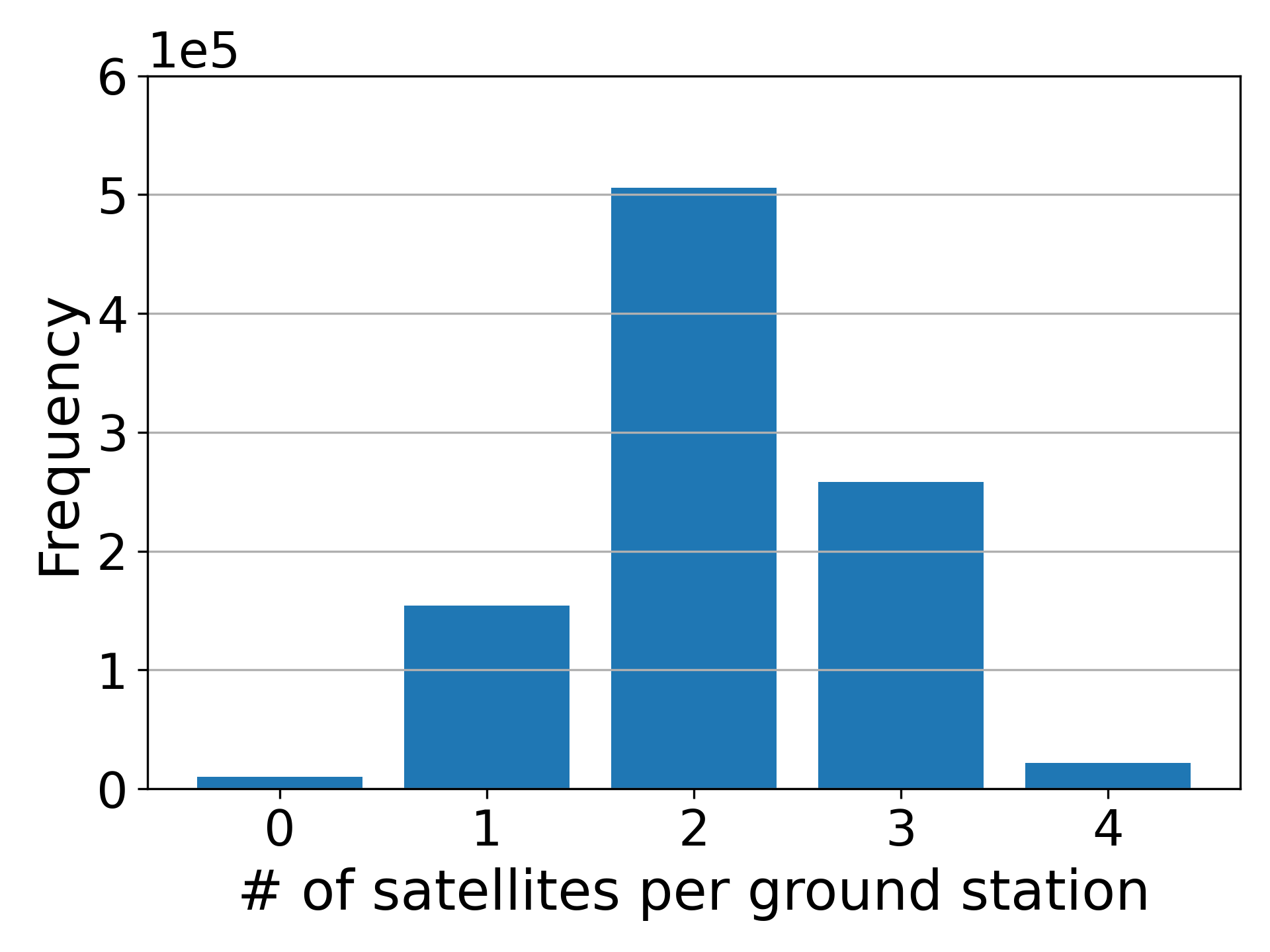}
        \caption{$A=500$ km, GS view.}
        \label{fig:Sat_Dist_500}
    \end{subfigure}
    % ---- Subfigure (b) ----
    \begin{subfigure}[b]{0.32\textwidth}
        \centering
        \includegraphics[width=\textwidth]{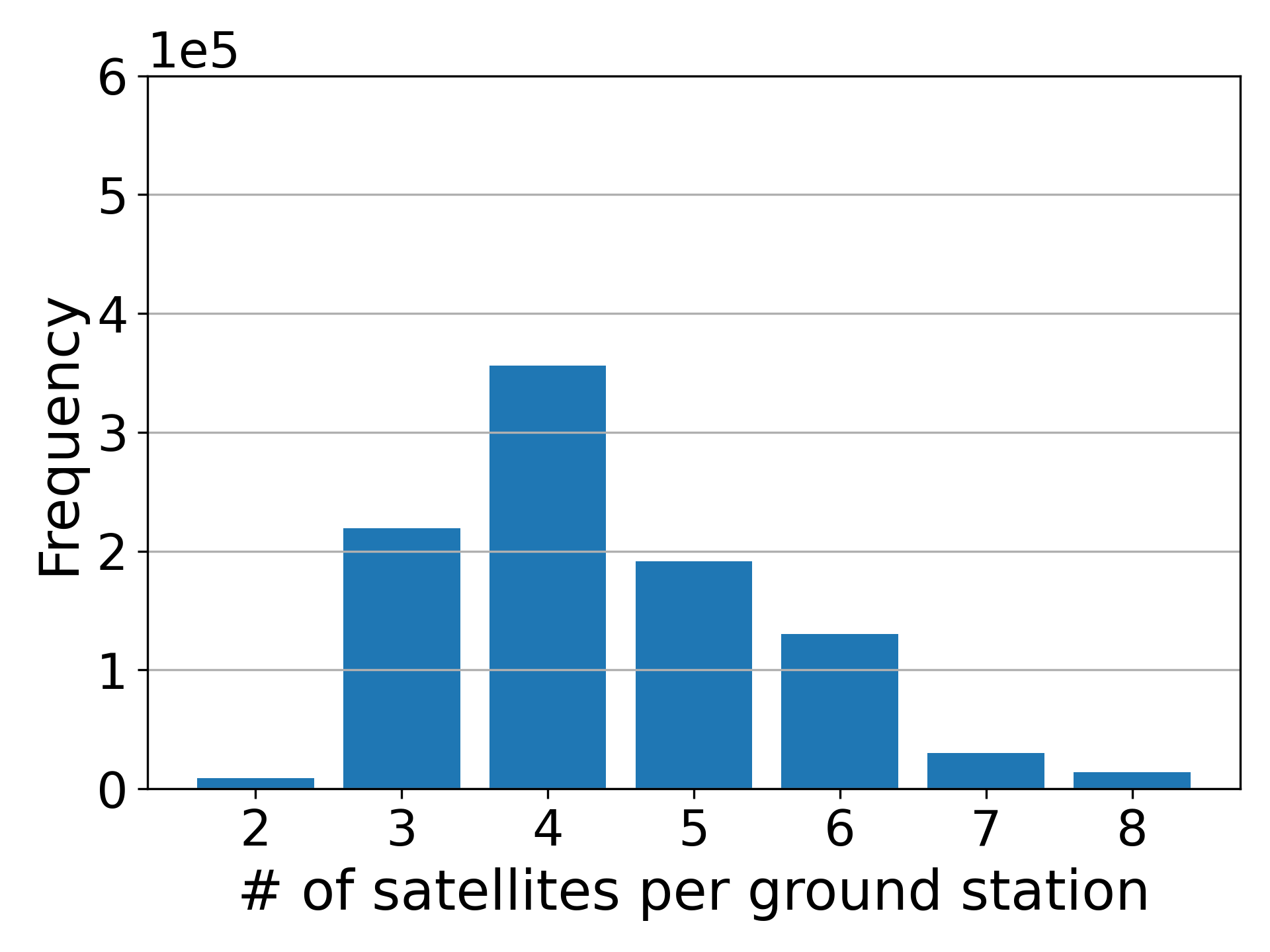}
        \caption{$A=800$ km, GS view.}
        \label{fig:Sat_Dist_800}
    \end{subfigure}
    % ---- Subfigure (c) ----
    \begin{subfigure}[b]{0.32\textwidth}
        \centering
        \includegraphics[width=\textwidth]{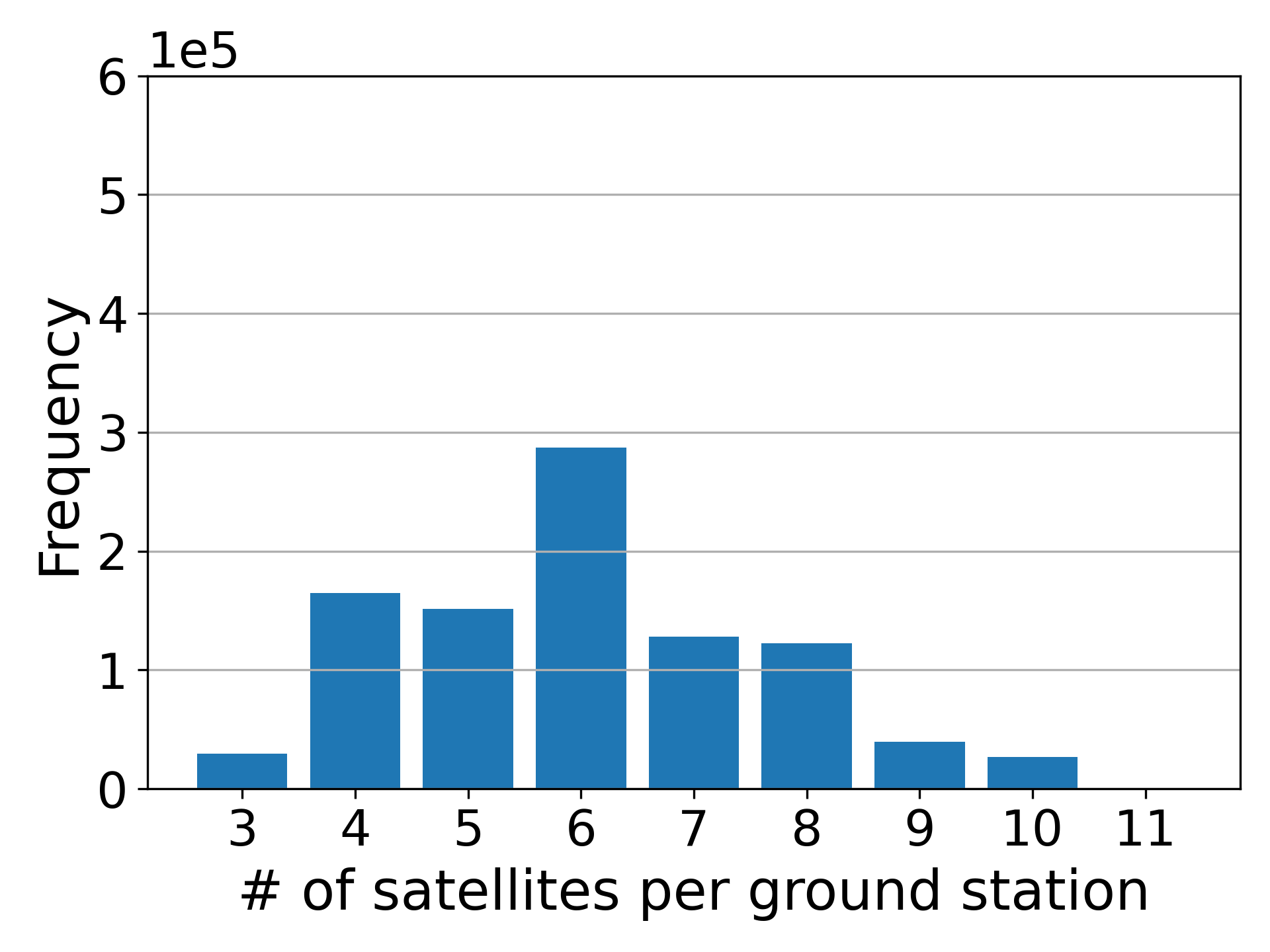}
        \caption{$A=1000$ km, GS view.}
        \label{fig:Sat_Dist_1000}
    \end{subfigure}
    \caption{Global QKD setting. Top row: satellite view, i.e., histogram of the number of ground stations that can be served by a satellite in a slot. 
    %considering all satellites and all the slots in a day. 
    Bottom row: ground station (GS) view, i.e., histogram of the number of satellites that can serve a ground station in a slot. Both rows consider all the slots for the day in September.
    }
    \label{fig:need-for-sche}
\end{figure*}

%% file: no-cloud.tex
%\subsection{Not Considering Weather Conditions}

\subsection{Results for Global QKD Setting}
We now present the results for the global QKD setting with 11 ground stations in 5 continents. We first present the results assuming no cloud coverage, i.e.,  $c_{t,g}=0$, $\forall t$, $\forall g \in \mathcal{G}$, in \S\ref{sec:no-cloud}. This is an idealistic setting that provides the best performance. It serves as an upper bound of realistic settings that have cloud coverage, which we present next in \S\ref{sec:res-with-cloud}.

\subsubsection{No Cloud Coverage} \label{sec:no-cloud}

%Show that minimum rate guarantee is indeed achieved. 400x11
\input{fig-phase-1}
\input{fig-pairwise-key}

%We now present key generation results across the ground station pairs by the various schemes. 

We first show the results for the two opportunistic schemes, Op-RR and Op-Greedy, during Phase 1. Specifically, these two schemes take respectively the key rate from RR and Greedy as input, and opportunistically schedule the satellites to serve the ground stations. 
%improve the overall system performance. 
Fig.~\ref{fig:phase-1}a shows a scatter plot of the total number of key bits generated for each  satellite and ground station pair over one day using Op-RR versus that obtained by RR, when $A=500$km and for the September day (the results for the other three days are similar). 
%\bing{Is there any zero values?(No) If so, did you omit them in the plot?} 
We see that the number of key bits obtained by Op-RR is significantly larger than the corresponding value under RR for most cases, except for those cases where these two schemes obtain the same results. Fig.~\ref{fig:phase-1}b compares the total number of key bits obtained for each satellite and ground station pair using Op-Greedy versus that obtained by Greedy. We again see that the opportunistic strategy leads to more or equal key bits for each satellite and ground station pair. Comparing Fig.~\ref{fig:phase-1}a and b, we see that the RR leads a much narrower range of values compared to the Greedy heuristic, which is consistent with the design goal of RR that aims to obtain similar number of key bits for satellite and ground station pairs.      

Fig.~\ref{fig:GRR_OS-matching} shows the key size (in bits) for each of the 55 ground station pairs for four schemes, RR, Greedy, Op-RR and Op-Greedy, sorted in descending order according to Op-RR. As expected, since Op-RR obtains more key bits per satellite and ground station pair than RR in Phase 1, the number of key bits per ground station pair under Op-RR is higher than that under RR. Similar argument holds for Op-Greedy versus Greedy. We see that the number of key bits per ground station pair is similar across the ground station pairs under the Greedy heuristic, which is 
%perhaps due to the averaging effect among the large 
not necessarily true in other scenarios, a point that we will return to later on. Last, while both Op-RR and Op-Greedy lead to more key bits than RR and Greedy, they lead to different distribution of key bits across the ground station pairs.

\begin{figure*}[t]
    \centering
    % ---- Subfigure (a) ----
    \begin{subfigure}[b]{0.32\textwidth}
        \centering
        \includegraphics[width=\textwidth]{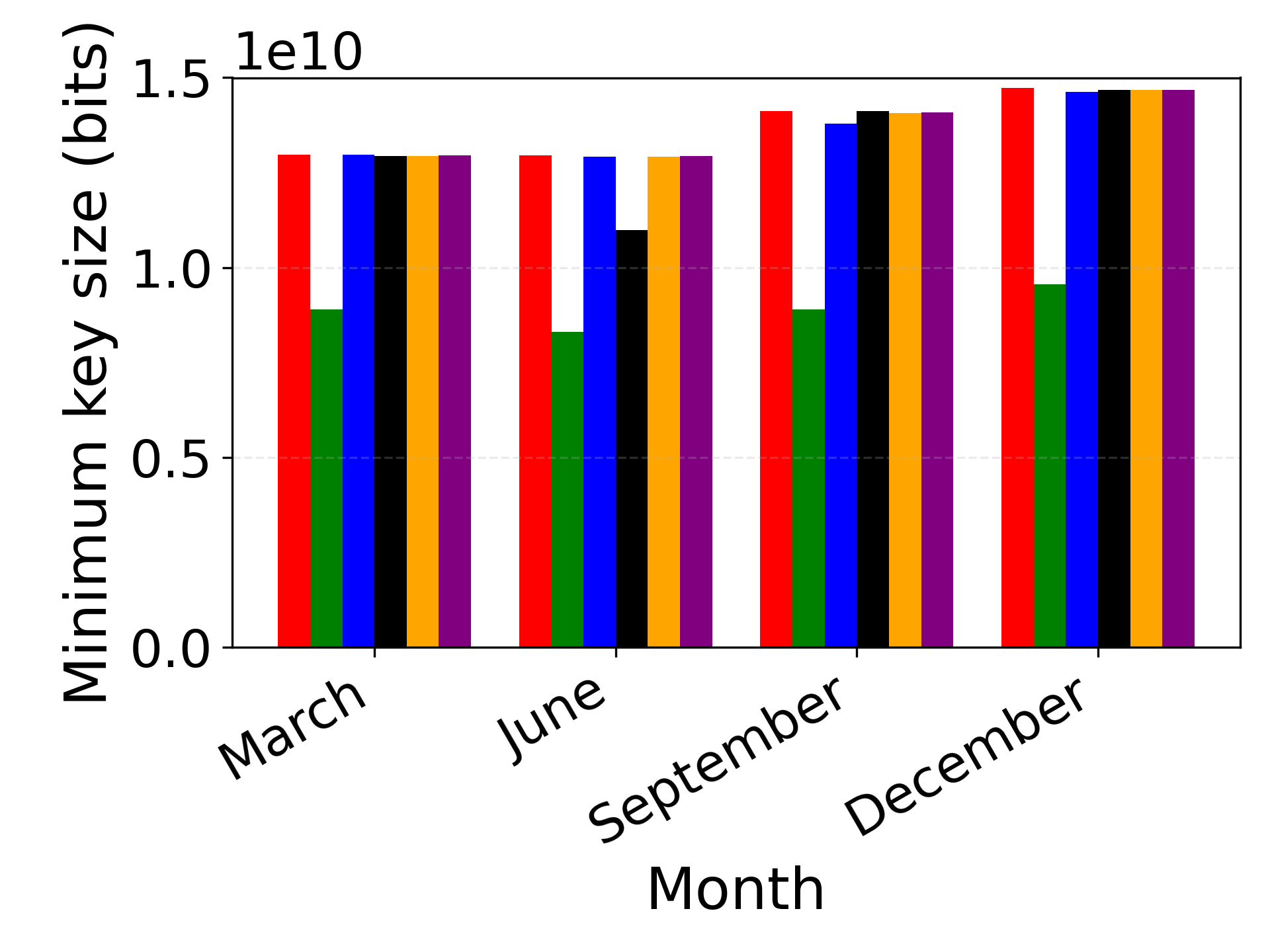}
        \caption{$A=500$ km.}
        \label{fig:minkey_500_WOCC}
    \end{subfigure}
    % ---- Subfigure (b) ----
    \begin{subfigure}[b]{0.32\textwidth}
        \centering
        \includegraphics[width=\textwidth]{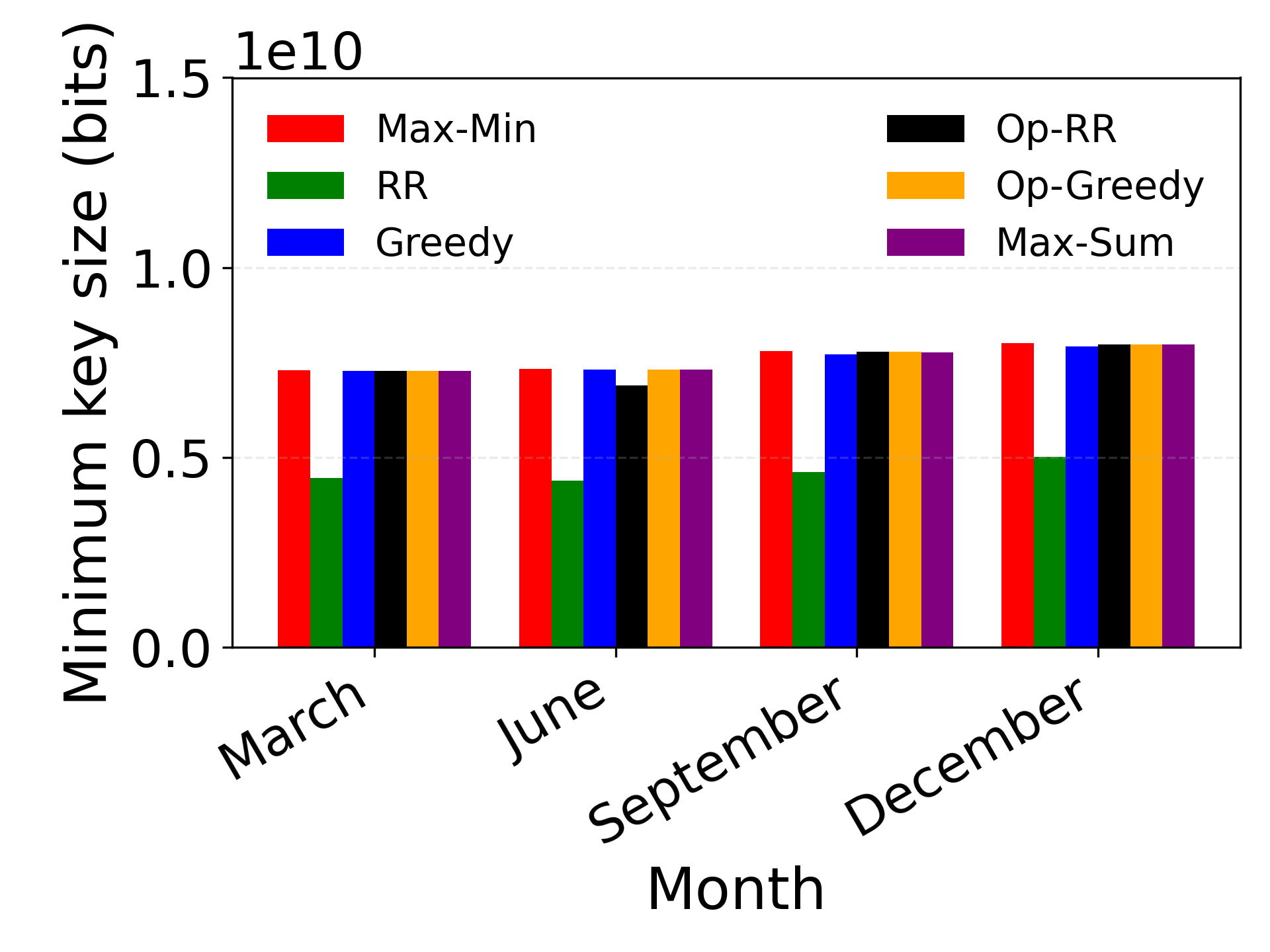}
        \caption{$A=800$ km.}
        \label{fig:minkey_800_WOCC}
    \end{subfigure}
    % ---- Subfigure (c) ----
    \begin{subfigure}[b]{0.32\textwidth}
        \centering
        \includegraphics[width=\textwidth]{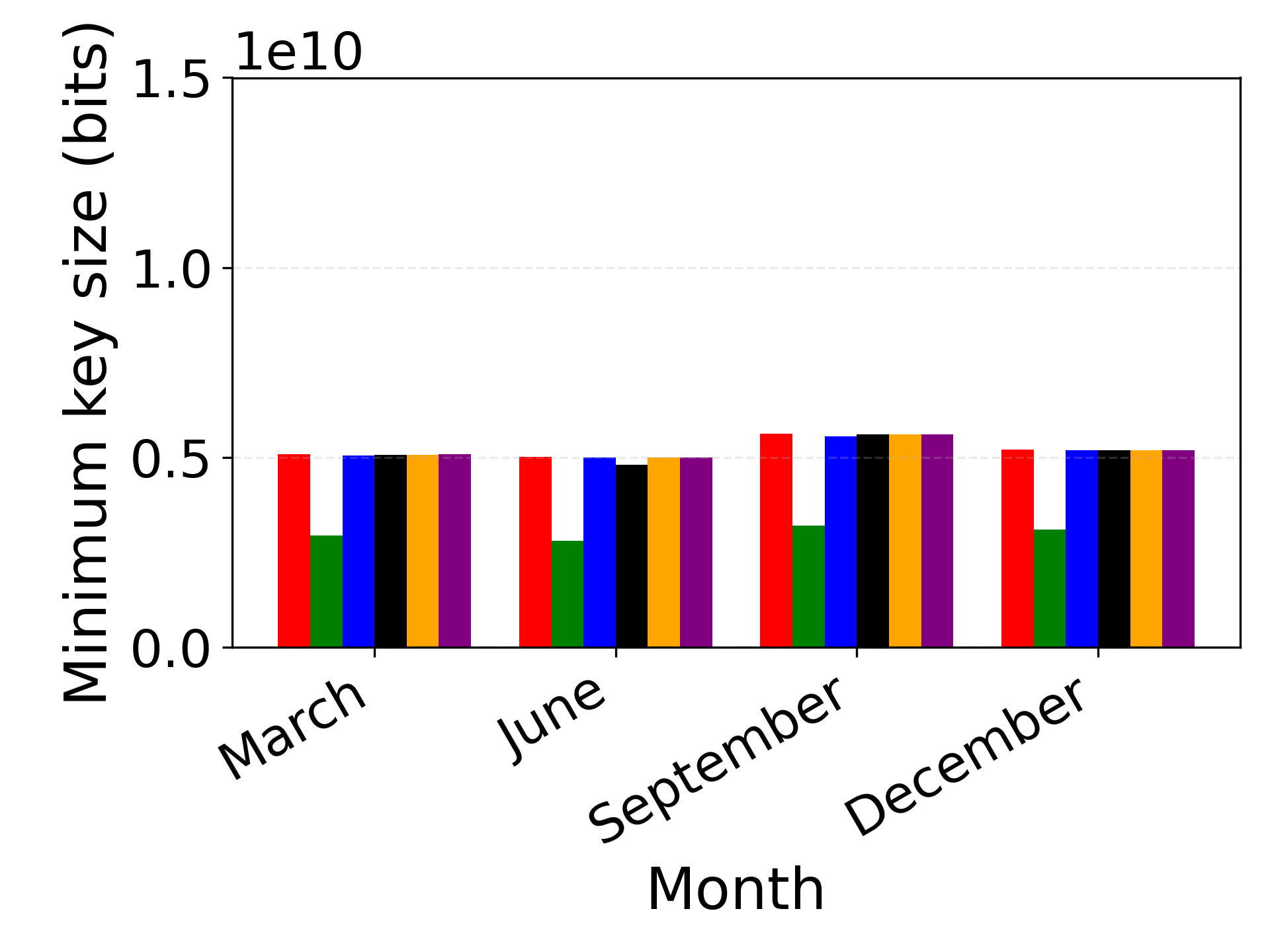}
        \caption{$A=1000$ km.}
        \label{fig:minkey_1000_WOCC}
    \end{subfigure}

    \caption{Minimum key size (bits) across the 55 ground station pairs in the global QKD setting, assuming no cloud coverage. 
    %\bing{y-axis: Minimum key size (bits). You can make the x and y caption slightly smaller.}
    %different seasons without cloud coverage: (a) 500, (b) 800, and (c) 1000. 
    %\bing{y-axis: Minimum \# of keys. Adjust the range of y-axis so that the bars are more differentiable? x-axis: Month, Use March, June, September, December instead of the seasons; March may not be called Spring in other places. Put the bars in the order of Max-Min, RR, Op-RR, Greedy, Op-Greedy, Max-Sum. Change the legend accordingly. }
    }
    \label{fig:minkey_WOCC}
\end{figure*}

Fig.~\ref{fig:minkey_WOCC} shows the minimum key size (in bits)  across all the ground station pairs. 
The results for 
Max-Min, RR, Greedy, Op-RR,  Op-Greedy, and Max-Sum are shown in the figure. 
%, the two opportunistic schemes,  their heuristic counterparts, as well as the results of Max-Min and Max-Sum.  
Max-Min leads to the optimal results for this setting. In contrast, Max-Sum favors the cases with high key rate, which can lead to unfair key rates to some ground station pairs.  Interestingly, for this setting, we see  Max-Sum leads to similar results as Max-Min in all the scenarios. This is perhaps due to the averaging effect when considering a large number of ground station pairs, a point that we will return to in \S\ref{sec:res-regional} for the regional QKD setting. RR leads to the lowest minimum key size in all the scenarios. Op-RR, on the other hand, 
leads to values close to the optimal value in most cases, except for the day in June and $A=500$ km. Greedy and Op-Greedy lead to results close to the optimal values in all the settings, which is not always true, as we shall see later when considering realistic settings with cloud coverage. 

In Fig.~\ref{fig:minkey_WOCC}, we see higher minimum key size for the two days in September and December than the other two days, which is particularly noticeable for $A=500$ km.  This is due to better weather conditions for most ground stations in these two days.
%compared to those for the  other two days.
For the same day, we see lower  minimum key size when increasing the satellite altitude. This is because, even though higher altitudes lead to better coverage, the transmissivity of the channel from a satellite to a ground station degrades with distance.  

%%[htbp]

\begin{figure*}[t]
    \centering
    % ---- Subfigure (a) ----
    \begin{subfigure}[b]{0.32\textwidth}
        \centering
        \includegraphics[width=\textwidth]{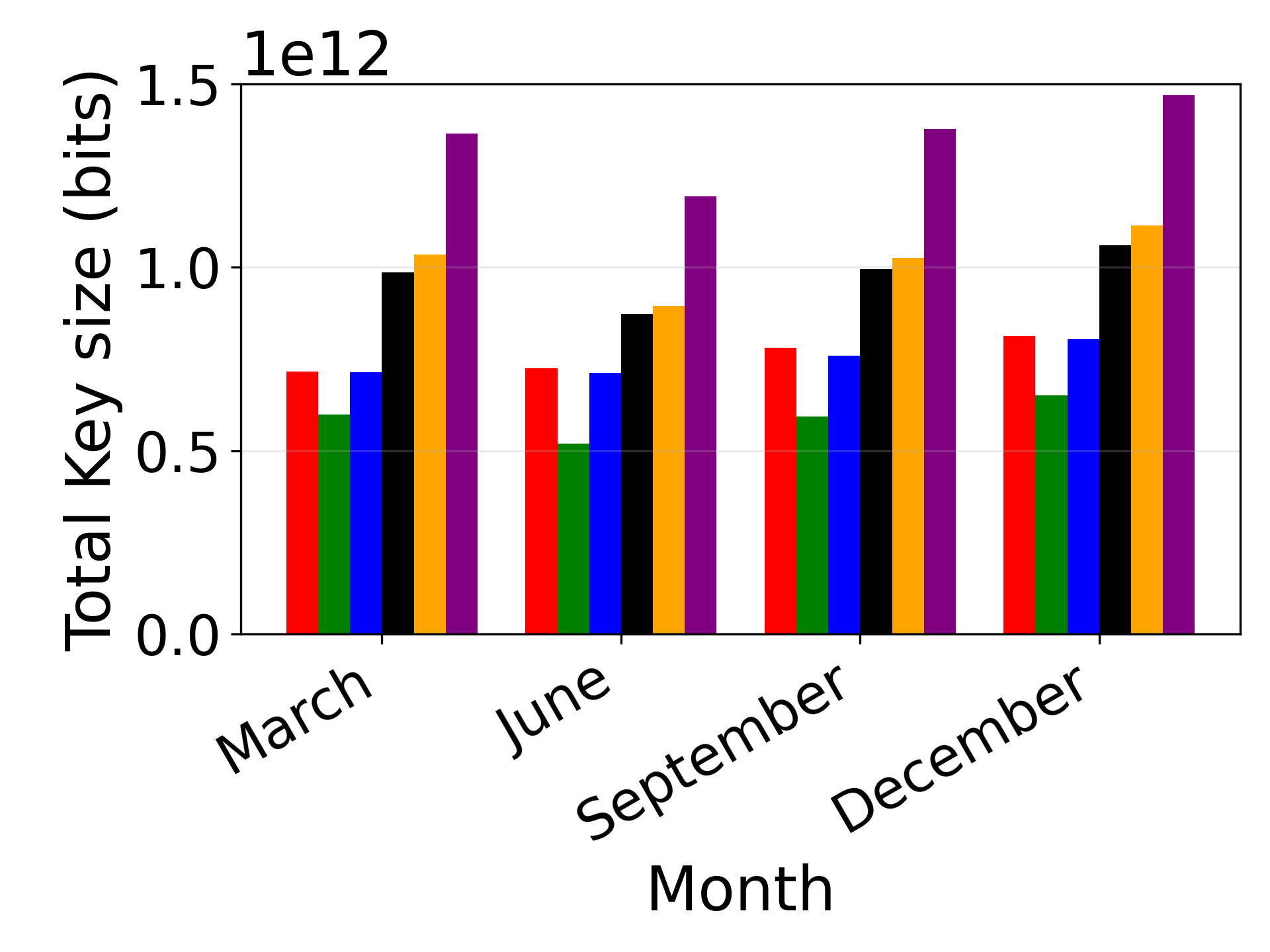}
        \caption{$A=500$ km.}
        \label{fig:totalkey_500_WOCC}
    \end{subfigure}
    % ---- Subfigure (b) ----
    \begin{subfigure}[b]{0.32\textwidth}
        \centering
        \includegraphics[width=\textwidth]{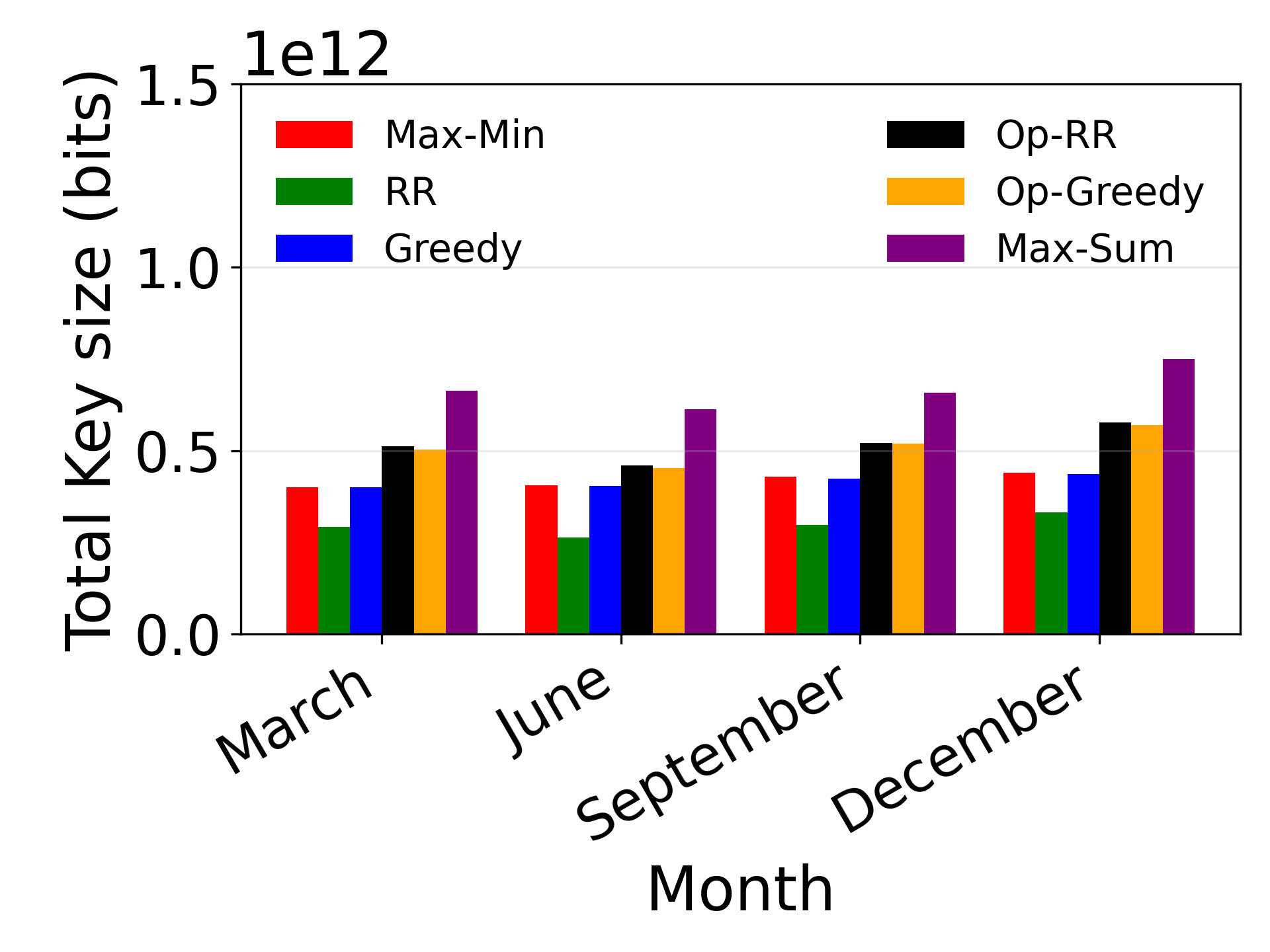}
        \caption{$A=800$ km.}
        \label{fig:totalkey_800_WOCC}
    \end{subfigure}
    % ---- Subfigure (c) ----
    \begin{subfigure}[b]{0.32\textwidth}
        \centering
        \includegraphics[width=\textwidth]{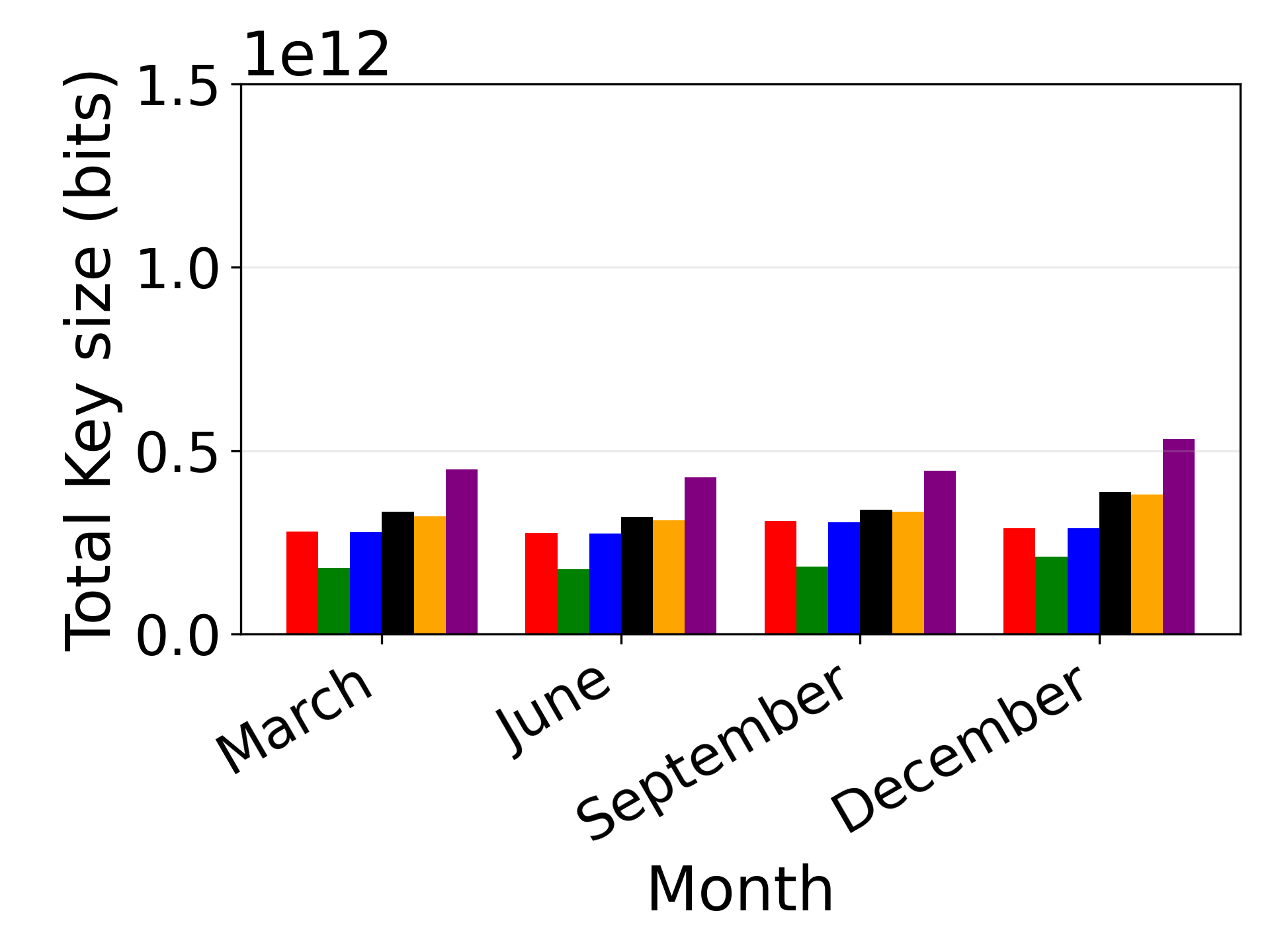}
        \caption{$A=1000$ km.}
        \label{fig:totalkey_1000_WOCC}
    \end{subfigure}

    \caption{Total key size (bits) over the 55 ground station pairs in the global QKD setting, assuming no cloud coverage.
}
    \label{fig:totalkey_WOCC}
\end{figure*}

%\bing{Show similar plots that compares the sum of keys across all the ground station pairs for all the schemes. }
Fig.~\ref{fig:totalkey_WOCC} plots the total key size (in bits) across the 55 ground station pairs, where Max-Sum leads to the optimal result in each scenario. The two opportunistic schemes lead to results closest to the optimal solutions in all the scenarios, significantly outperform their heuristic counterparts, and the results from Max-Min. 
Specifically, the value of Op-RR is 72\% to 73\% of the corresponding optimal values, while for Op-Greedy, the range is 75\% to 76\%. 
For each altitude, the results for the four seasons are similar. We see lower total key size when increasing the satellite altitude due to lower transmissivity.

%% file: fig-phase-1.tex
\begin{figure*}[t]
    \centering
    % ---- Subfigure (a) ----
    \begin{subfigure}[b]{0.40\textwidth}
        \centering
        \includegraphics[width=\textwidth]{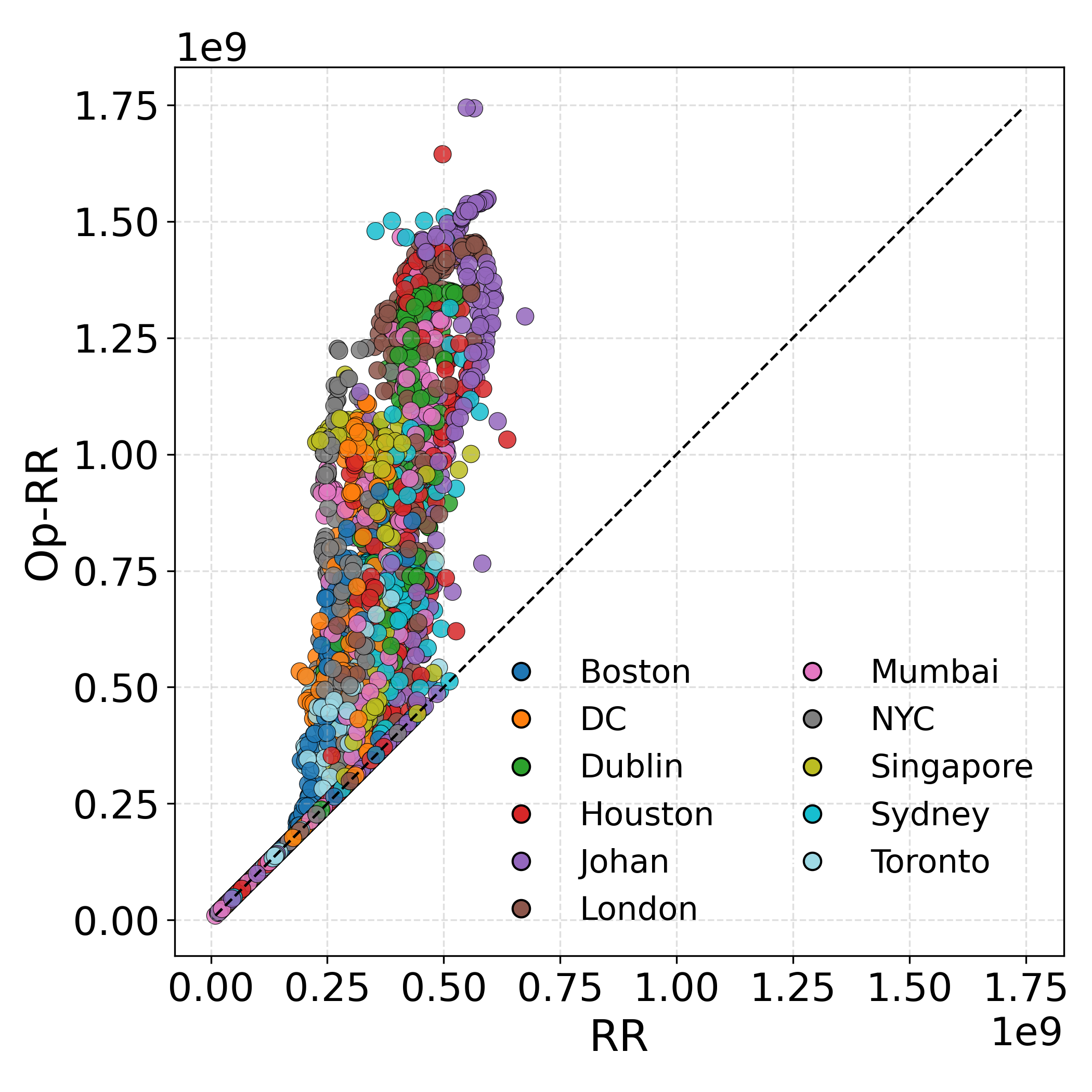}
        \caption{Op-RR versus RR.}
        \label{fig:Op-RR-RR}
    \end{subfigure}
    % ---- Subfigure (b) ----
    \begin{subfigure}[b]{0.40\textwidth}
        \centering
        \includegraphics[width=\textwidth]{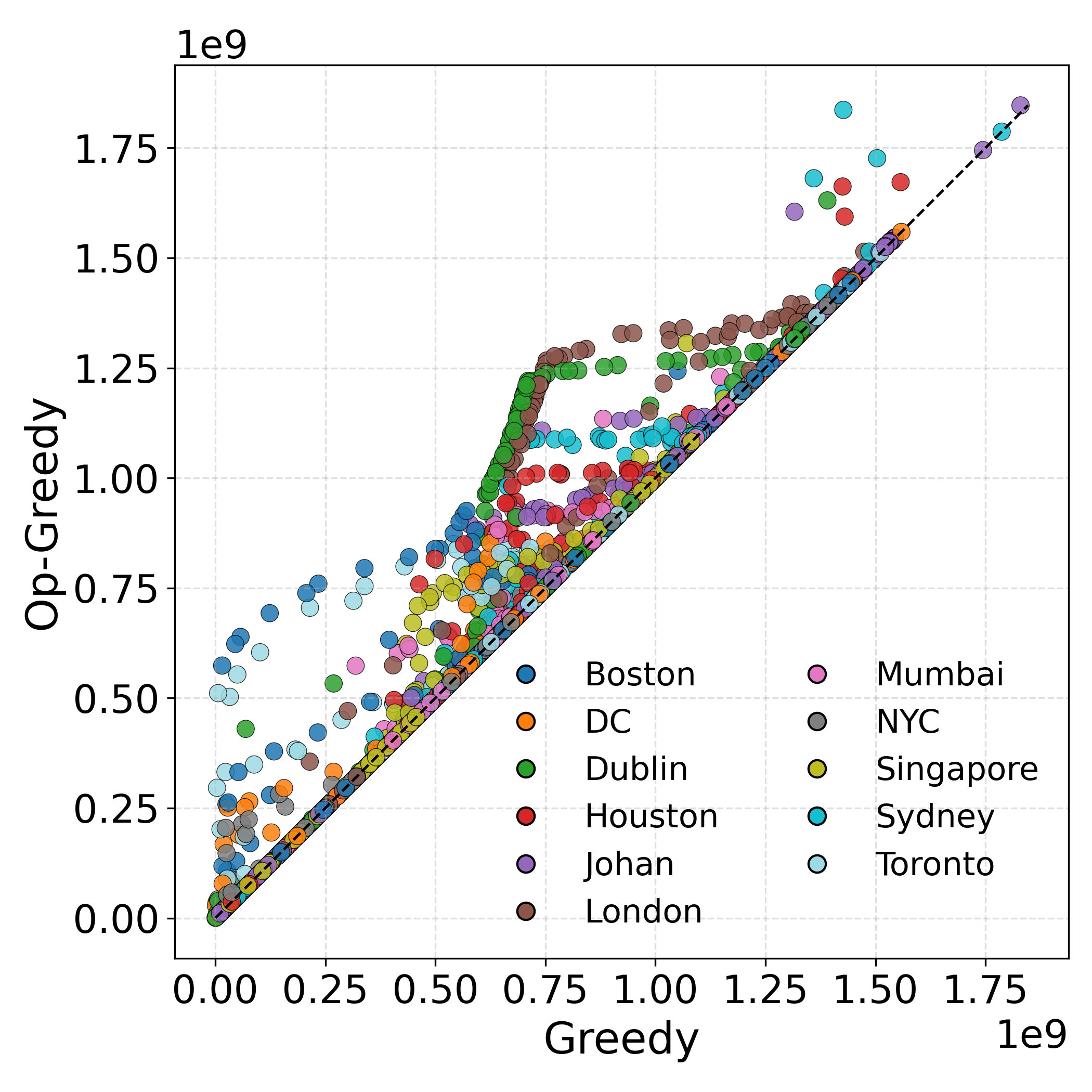}
        \caption{Op-Greedy versus Greedy.}
        \label{fig:Op-Greedy-Greedy}
    \end{subfigure}

    \caption{Key size (in bits) obtained for each satellite and ground station pair at the end of one day (i.e., Phase 1 results), global QKD setting, $A=500$km, for the day in September.
    }
    \label{fig:phase-1}
\end{figure*}

%% file: fig-pairwise-key.tex
\begin{figure}[h]
\centering
\includegraphics[width=0.9\textwidth]{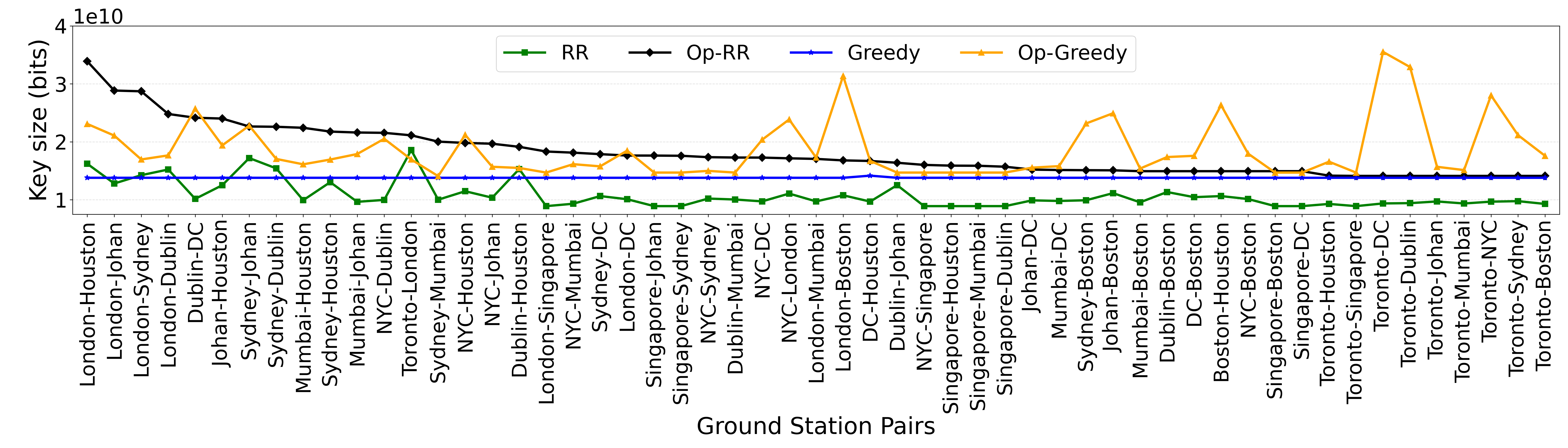}
\caption{Key size (bits) for each of the 55 ground station pairs in the global QKD setting for Op-RR and Op-Greedy versus RR and Greedy, $A=500$ km, for the day in September, assuming no cloud coverage. 
}
\label{fig:GRR_OS-matching}
\end{figure}

%% file: with-cloud.tex
\subsubsection{Considering Cloud Coverage} \label{sec:res-with-cloud}

\input{fig-cloud-phase-1}

%maybe compare two ground stations, one with high cloud coverage and the other with low clound coverage; show the results for Phase 1, and then pairwise results 10 + 10; motivate filtering some ground stations on an hourly basis, show that the filtering leads to even higher sum of keys 

We now present the results when considering cloud coverage. For each day, for convenience, we obtain the cloud coverage at the beginning of each hour (see \S\ref{sec:model}) and use it as the cloud coverage for each slot in that hour.  Fig.~\ref{fig:cloud} plots the average cloud coverage for each ground station for the four days in March, June, September, and December, sorted in descending order according to the cloud coverage in March; the standard deviation is large for many cases and is omitted in the plot for clarity. For each day, we see significantly different cloud coverage across the ground stations. For the same ground station, the cloud coverage can also differ substantially since the four days correspond to different seasons. For the day in June, all the ground stations have generally low cloud coverage,
%than the other three days, 
while DC has the highest average cloud coverage in the day in December, followed by Sydney for the day in September.

Since some ground stations have high cloud coverage in some slots, which leads to low key rate, including them in the scheduling problem can lead to inefficient usage of the resources. %since the key rate to such ground stations is low due to cloud coverage. 
%, while the key rate to consumes the resources of the system, while  for those slots can lower the number of key bits to other ground stations,
%that are in more favorable conditions, 
%and hence reduce the overall number of keys in the system. 
We therefore consider a {\em filtering} approach, which ignores a ground station in the scheduling problem in a slot if its cloud coverage in that slot is above a threshold.  In the rest of the paper, we use a threshold of 0.8. 
%Since our cloud coverage data is recorded per hour, this approach leads to filtering at the 
%We next compare the results with and without filtering. 
When using the above filtering approach, if a ground station has cloud coverage above the threshold for an entire day, it will not obtain any key with any other ground stations. For the days in March and December, one ground station (Boston and DC, respectively) ends up having no keys with others. For the day in September, two ground stations (Sydney and Mumbai) have no keys with others. For the day in June, all ground stations have positive key rate with others.      

%wb, 10/10/2025, talk about minimum and total key size together: log scale since some key sizes are an order of magnititude lower. (i) both are significantly lower than without cloud, June best with cloud since all the ground stations have generally low cloud coverage; different from the no cloud case.  (ii) Go through the solid bars first. (iii) talk about the difference between solid and dashed bars. min key size: relative performance of the different schemes (with and without filtering); what filtering does to max-min (mostly reduce the key size, except for one case).   

\input{fig-cloud-res}

The top and bottom plots in  Fig.~\ref{fig:key_cloud} show respectively the minimum and total key size  across the ground station pairs for $A=500$km; the results for the other two altitudes show similar trends and are omitted.
In the figure, the solid color bars represent the results when not applying filtering, while the bars with slashed lines represent the results when applying filtering. We use log scale for these two plots since the values in each plot can differ in orders of magnitude.

We first present the results when not applying filtering (i.e., the solid color bars). 
As expected, due to cloud coverage, 
%for each scheme and setting, 
both minimum and total key sizes are significantly lower than those without cloud coverage (see  Fig.~\ref{fig:minkey_WOCC}a and Fig.~\ref{fig:totalkey_WOCC}a). Specifically, 
for minimum key size, the largest value is $3.4\times 10^9$ bits (Max-Min for the day in June), only 23\% of the largest value without cloud coverage.
%(Max-Min for the day in September). 
For total key size, the largest value is $8.5\times 10^{11}$ bits (Max-Sum for the day in June), only 58\% of the largest value without cloud coverage. The number of key bits for the day in June is larger than those for the other days since the cloud coverage for that day is low for all ground stations. For minimum key size, we again see that the two opportunistic schemes and Max-Sum lead to results close to the optimal value (obtained by Max-Min) for all four days. For total key size, the two opportunistic schemes are closer to the optimal values (obtained by Max-Sum) than other schemes. Specifically, the  total key size obtained by Op-RR is 63\%-76\% of the optimal value for the four days, while the range is  68\%-82\% for Op-Greedy.    

We next present the results when applying filtering (i.e., the bars with slashed lines). For minimum key size, the ground stations pairs with zero key are excluded from the plot (since  all schemes have minimum key size as zero in that case). Op-RR is close to the optimal value (by Max-Min) for all the four days, while Op-Greedy is only close to the optimal value for one day (in June); for the other three days, its value is only 18\%, 37\%, and 68\% of the optimal value. For total key size, Op-RR again leads to results close to the optimal values (by Max-Min),
%for all the four days, 
while Op-Greedy leads to results only 31\% to 42\% of the optimal values. Therefore,
%above results demonstrate that 
Op-RR, which starts with a fairer key 
%rate 
allocation by RR than Greedy, tends to lead to better results than Op-Greedy. Indeed, 
%Op-Greedy is based on the Greedy heuristic, which 
Greedy allocates satellites in a greedy manner in Phase 1, which may cause a satellite to have unbalanced number of key bits with ground stations, and hence does not help ground station pairs to generate keys through this satellite in Phase 2. Op-Greedy, based on Greedy, may have the same limitation. 

%a lot of keys with one ground station, and few keys with another ground station, which does not help these two ground station does not help satellites to share keys with ground statican cause lead to detrimental impact in which starts with a key rates that are less fair than RR. 

%wb, 10/13/2025, modified version
We now compare the results when applying filtering with those when not applying filtering (i.e., comparing two adjacent bars, one solid color and the other with slashed lines in Fig.~\ref{fig:key_cloud}). For minimum key size, we only compare the results for the day in June since for all the other three days, the key rate for some ground station becomes zero due to filtering. We see for all the schemes, using filtering leads to lower minimum key size. This is as expected since the ground stations with high cloud coverage are only served in the slots with cloud coverage lower than the threshold, which can further limit their key sizes and lower the minimum key size. 
For total key size, filtering leads to different impacts for different schemes. We next describe its impacts on Op-RR, which is efficient
and provides the best tradeoff of all the schemes. When using filtering, the total key size under Op-RR is 2.5-2.9$\times$ of that without filtering, demonstrating the benefits of using filtering in improving system throughput.

%% file: fig-cloud-phase-1.tex
\begin{figure}[t]%left, bottom, right, top
   \centerline{\includegraphics[width=0.42\textwidth, trim = 0.0cm 0cm 0.0cm 0cm, clip]{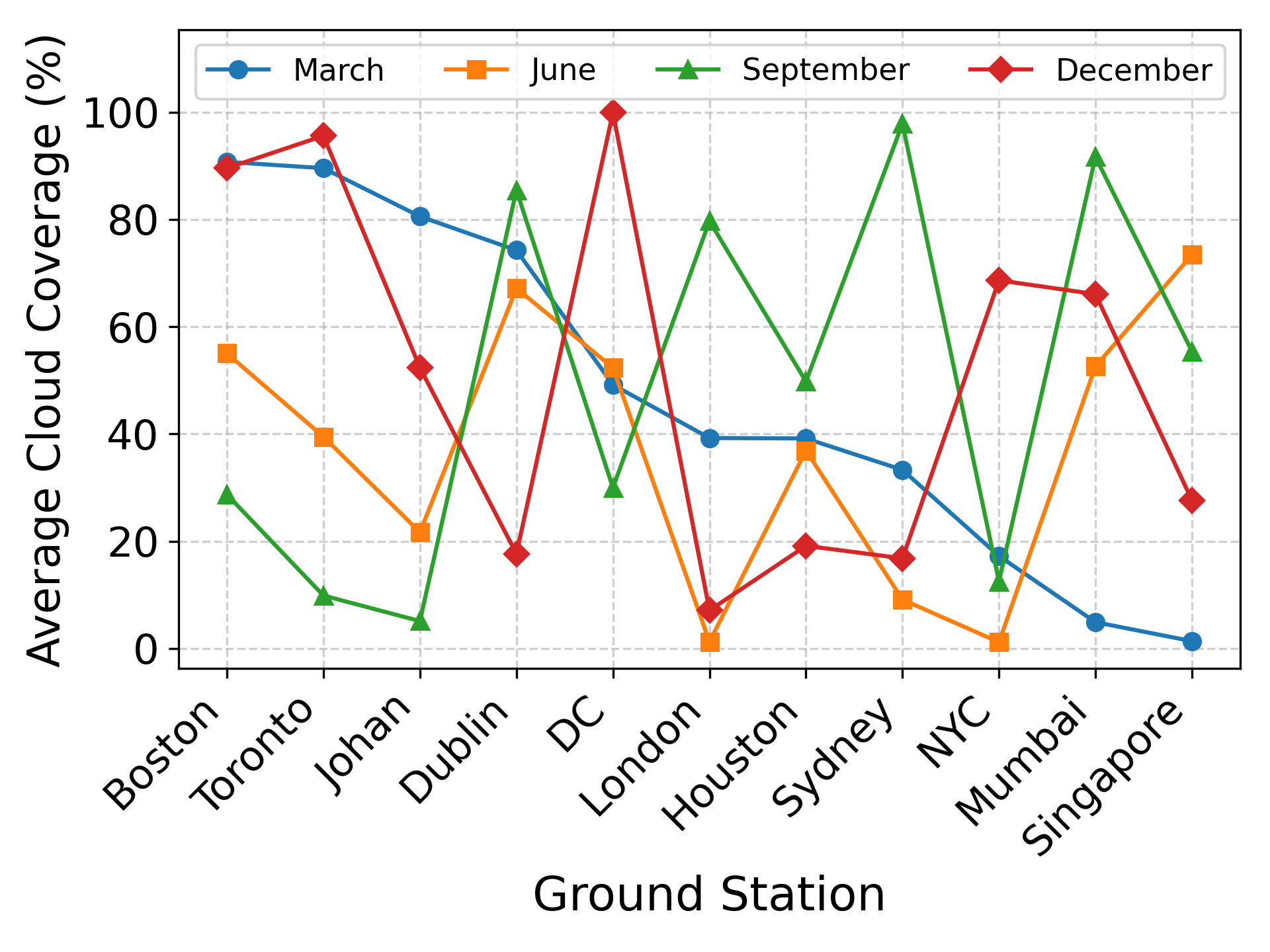}}
   \vspace{-0.15in}
\caption{{Average cloud coverage over a day for the 11 ground stations; standard deviation omitted for clarity.} 
}
\label{fig:cloud}
\vspace{-0.10in}
\end{figure}

%% file: fig-cloud-res.tex
\begin{figure*}[t]
    \centering
    \vspace{-0.1in}
    % ---- Subfigure (a) ----
    \begin{subfigure}[b]{0.90\textwidth}
        \centering
        \includegraphics[width=\textwidth]{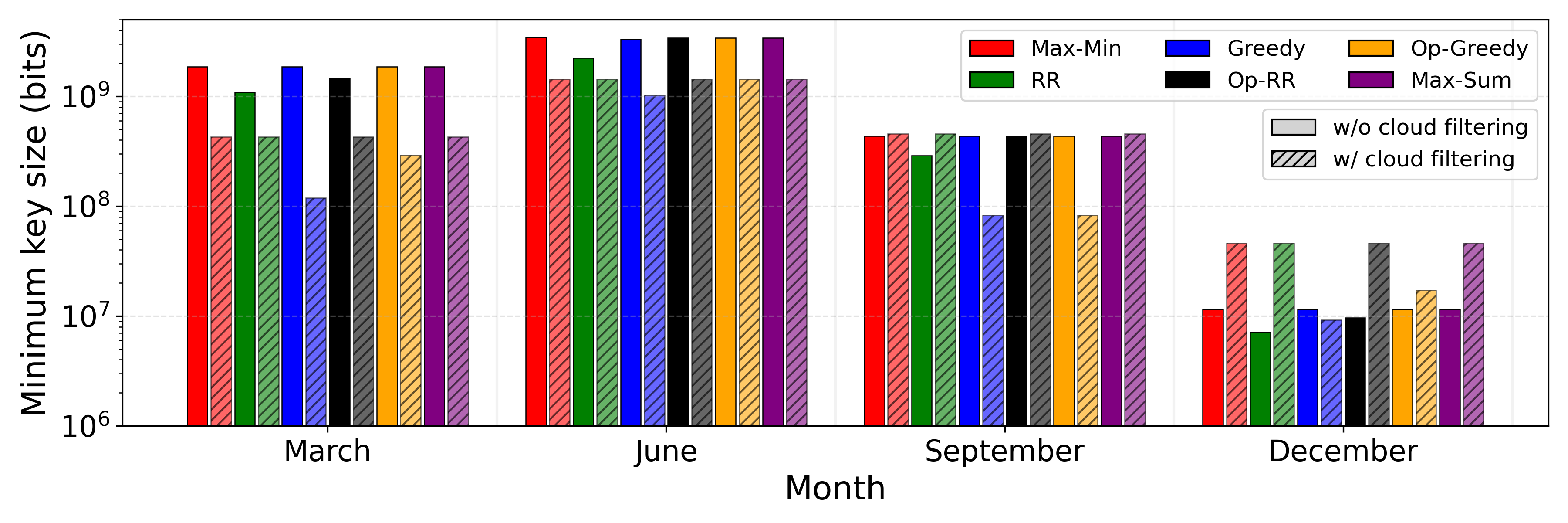}
        %\caption{Minimuvsm key size (bits).}
        \label{fig:minkey_500_WOCF_WCF}
    \end{subfigure}
    \vspace{-0.2in}
    %\vspace{0.5in}
    \begin{subfigure}[b]{0.90\textwidth}
        \centering
        \includegraphics[width=\textwidth]{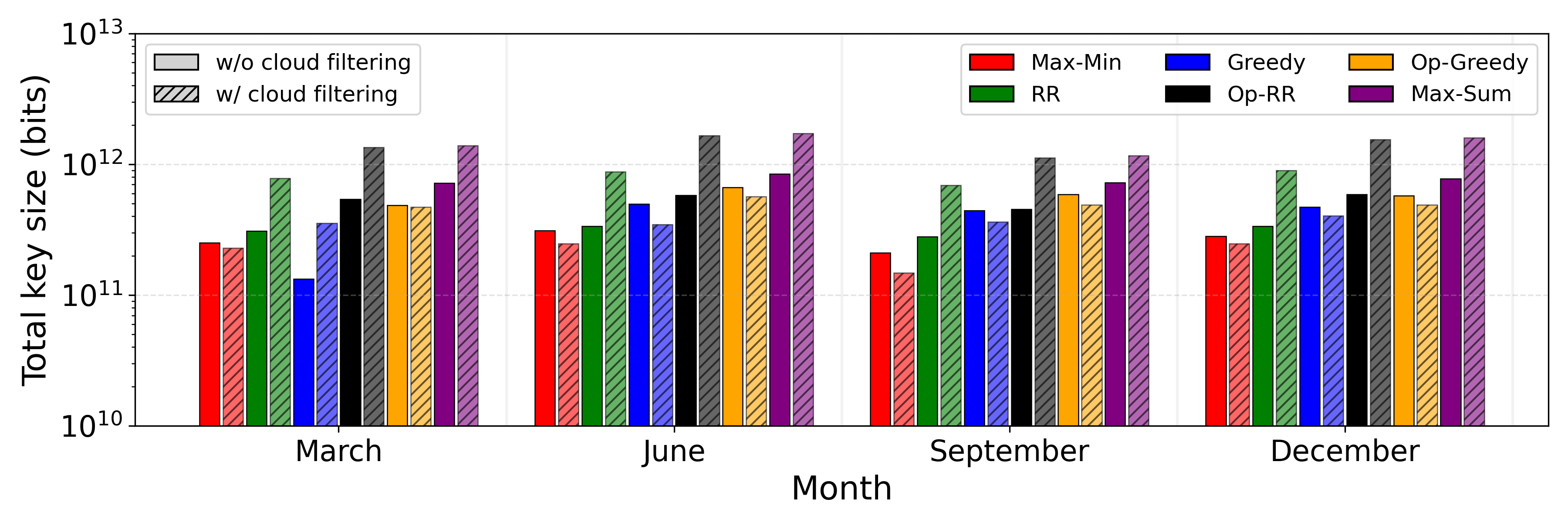}
        %wb, 10/10/2025, not show caption
        %\caption{Total key size (bits).}
        \label{fig:totalkey_500_WOCF_WCF}
    \end{subfigure}
    %\vspace{0.15in}
    \caption{Minimum (top) and total (bottom) key sizes of the ground station pairs, global QKD setting, $A=500$ km, considering cloud coverage. The results with and without filtering are represented as solid color bars and bars with slashed lines. Key size can be zero for some ground station pairs due to filtering, which is excluded in obtaining the minimum key size in the top plot. Note the different y-axis values in the two plots.
    %top plot are significantly lower than those in the bottom plot.
    }
    \label{fig:key_cloud}
\end{figure*}    
%\fi 

%% file: small-set-res.tex
\subsection{Results for Regional QKD Setting} \label{sec:res-regional}

%%ground stations filtered 
%March: 
%June: 
%September: 
%December: 

%So far, we have considered 11 ground stations in 5 continents. 
%We now present the results 
For the regional QKD setting (i.e., 4 ground stations in North America), 
%As for the global QKD setting, 
we again consider three scenarios: assuming no cloud coverage, considering cloud coverage without and with filtering. 
%
%when assuming no cloud coverage (which provides an upper bound), considering cloud coverage with no filtering, and considering cloud coverage with filtering (i.e., excluding a ground station in a slot if its cloud coverage is above 0.8). 
When using filtering, DC is not served for the day in December since its cloud coverage is above the threshold for all the slots. 
%
%Therefore, the results for this setting below are for 3 (instead of 6) ground station pairs.    

%When applying filtering, the number of ground stations \bing{drops to xx in xx, xx in xx}; for the other days, the number of ground stations remains to be 4.

\input{fig-4GS}

%We first present the minimum key size results. 
Fig.~\ref{fig:minkey_4gs-500km} shows the minimum key size when $A=500$ km for the three scenarios; the results for $A=800$ and 1000 km show similar trends and are omitted. The two opportunistic schemes lead to close to optimal values (obtained using Max-Min), similar to the observation for the global QKD setting. %We see that the two opportunistic schemes lead to close to optimal results (obtained using Max-Min). 
Max-Sum leads to significantly worse performance in most cases. For instance, it leads to minimum key size only 42\%-49\% of the optimal values when assuming no cloud coverage.
%wb, 10/13/2025, too many details
%
%10\% to 36\% of the optimal values when considering cloud coverage and no filtering, and 21\% to 81\% of the optimal values when considering cloud coverage and using filtering. 
%except for one case (considering cloud coverage and using filtering for the day in December). 
This is in contrast to the observations in the global QKD setting, where Max-Sum leads to close to optimal results. This difference is perhaps due to the small number of ground stations in the regional QKD setting, which lacks the natural averaging effect when there are a large number of ground stations at geographically distributed locations. 
As the global setting, we see that using filtering reduces minimum key size compared to not using filtering (see Figures~\ref{fig:minkey_4gs-500km}b and c, excluding the day in December since DC has no key for that day due to filtering).

Fig.~\ref{fig:totalkey_4gs-500km} shows the total key size (in bits) for the three scenarios when $A=500$ km. We see that the results under Op-RR is close to the optimal solution (obtained using Max-Sum): it is 88\%-91\%, 82\%-95\% and 82\%-91\% of the optimal values
for the three settings in Figures~\ref{fig:totalkey_4gs-500km}a to c. 
%when not assuming cloud coverage, 82\%-95\% of the optimal values when considering cloud coverage and no filtering, and 82\%-91\% of the optimal values when considering cloud coverage and no filtering. 
The gap between Op-RR and Max-Sum is even smaller in this setting than that in  the global QKD setting. Op-Greedy leads to results close to the optimal solution, except for the December day in Fig.~\ref{fig:totalkey_4gs-500km}c, where its minimum key size is 65\% of the optimal value. 
%
%Last, we compare the total key size when considering cloud coverage, with and without filtering in Figures~\ref{fig:totalkey_4gs-500km}b and c. 
Unlike the global QKD setting, we see that applying filtering reduces the total key size for Max-Sum, Op-RR and Op-Greedy for all of the four days (see Figures~\ref{fig:totalkey_4gs-500km}b and c). This might be due to the small number of ground stations in a small geographic region, where the resource is abundant, and hence not serving one ground station does not lead to much benefits to other ground stations.

%% file: fig-4GS.tex
\begin{figure*}[t]
    \centering
    %no cloud
    % ---- Subfigure (a) ----
    \begin{subfigure}[b]{0.32\textwidth}
        \centering
        \includegraphics[width=\textwidth]{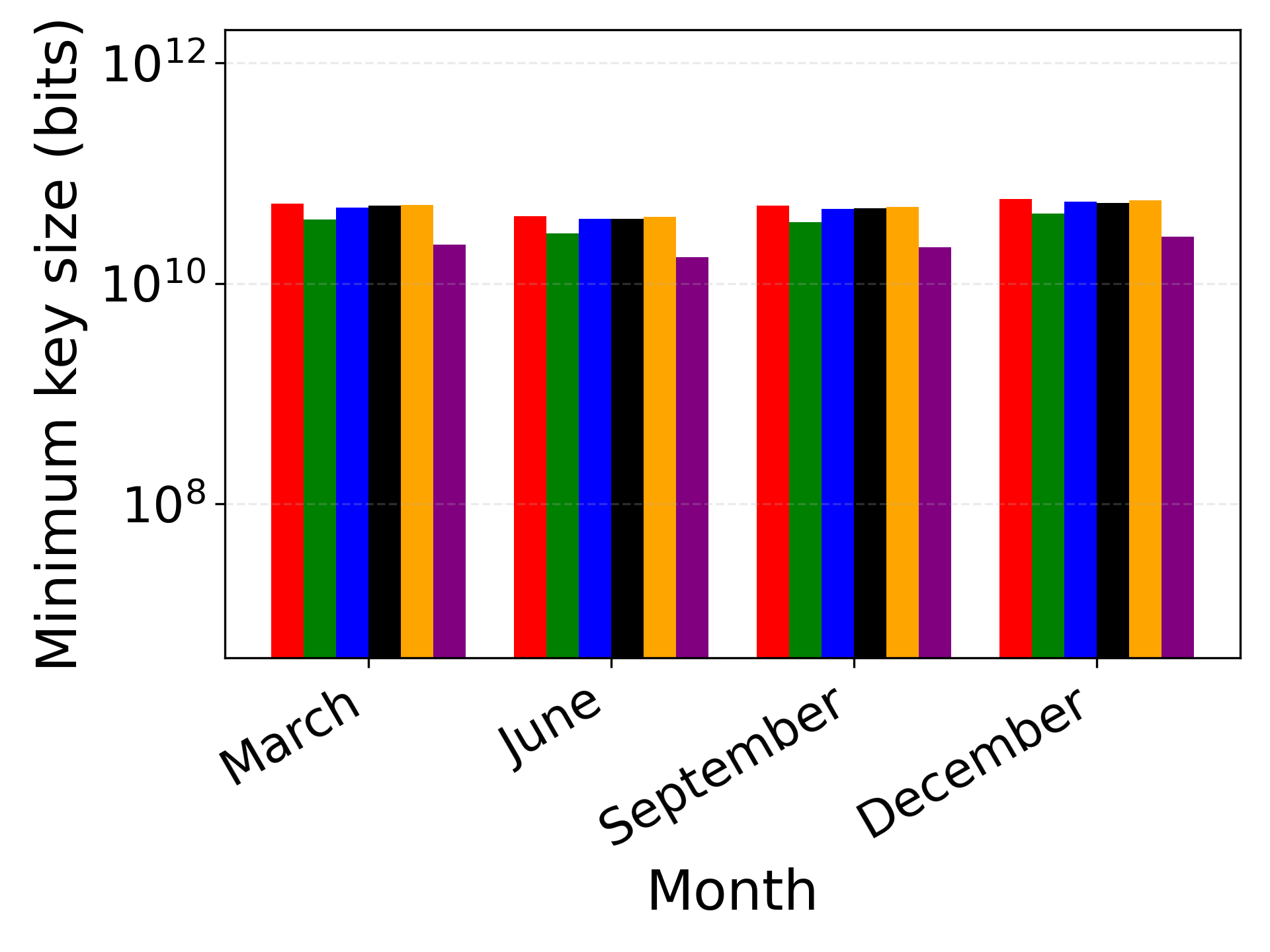}
        \caption{Assuming no cloud coverage.}
        \label{fig:minkey_4s_500_WOCC}
    \end{subfigure}
%with cloud, no filtering
        \begin{subfigure}[b]{0.32\textwidth}
        \centering
        \includegraphics[width=\textwidth]{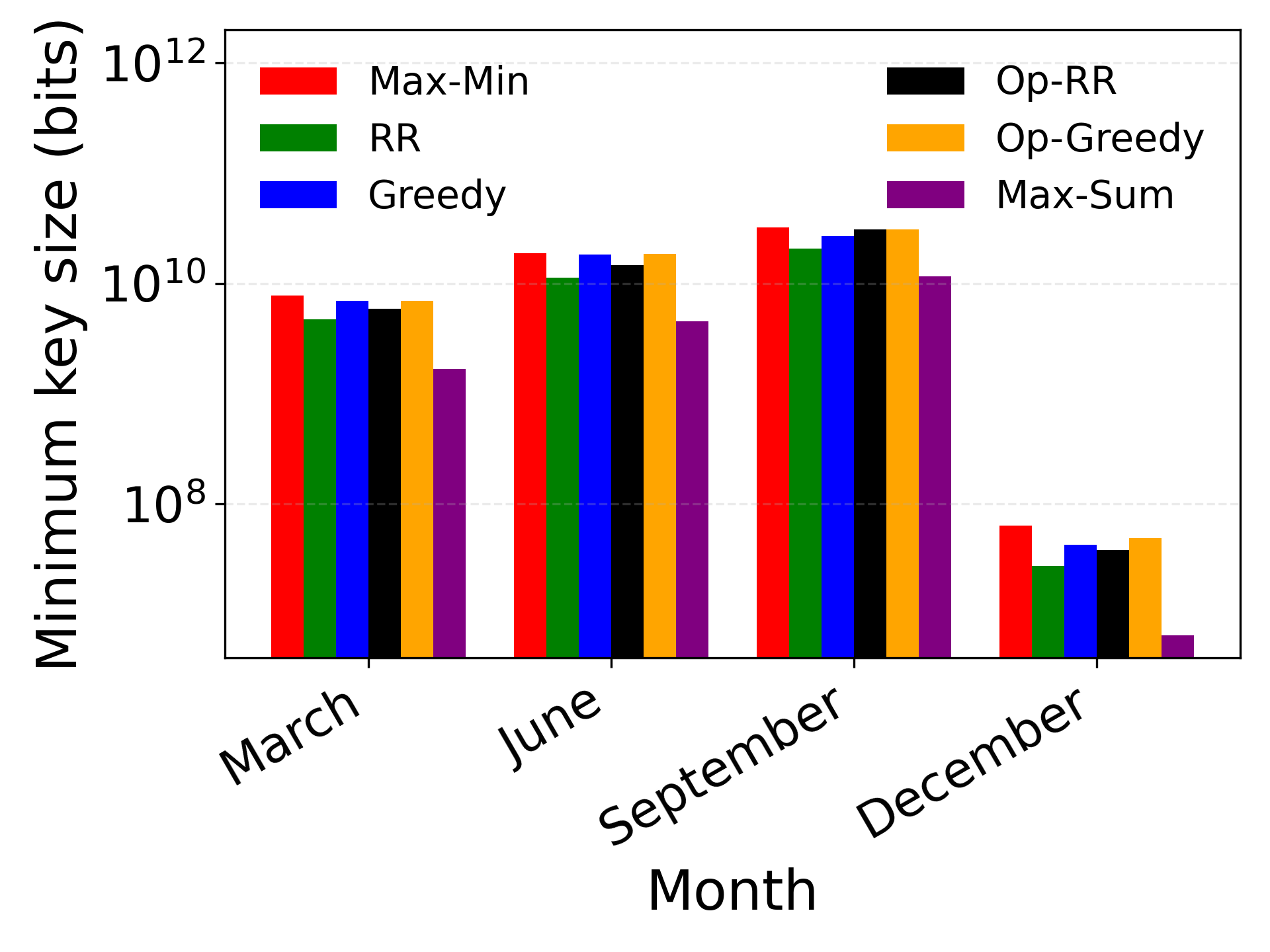}
        \caption{W/ cloud coverage, no filtering.}
        \label{fig:minkey_4gs_500_WCC}
    \end{subfigure}
    %with cloud, with filtering
    \begin{subfigure}[b]{0.32\textwidth}
        \centering
        \includegraphics[width=\textwidth]{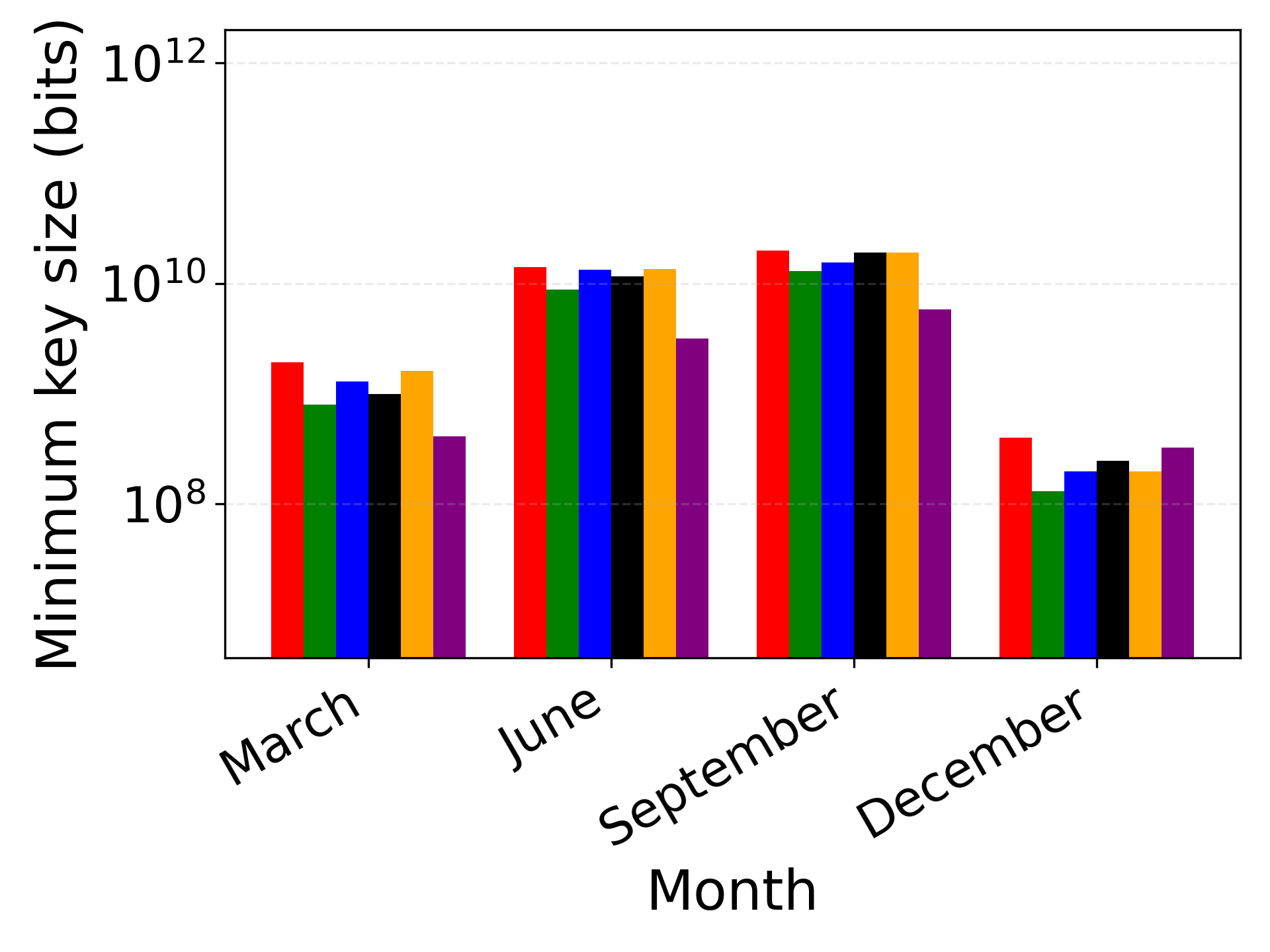}
        \caption{W/ cloud coverage, w/ filtering.}
        \label{fig:minkey_4s_500_CF}
    \end{subfigure}
    \caption{Minimum key size (bits) for the regional QKD setting, $A=500$ km. In plot (c), DC has zero key with any other ground station due to filtering in December, and is excluded from the plot.
    }
    \label{fig:minkey_4gs-500km}
\end{figure*}    

%total size results, only show the results for 500 km
\begin{figure*}[t]
    \centering
    %no cloud
    % ---- Subfigure (a) ----
    \begin{subfigure}[b]{0.32\textwidth}
        \centering
        \includegraphics[width=\textwidth]{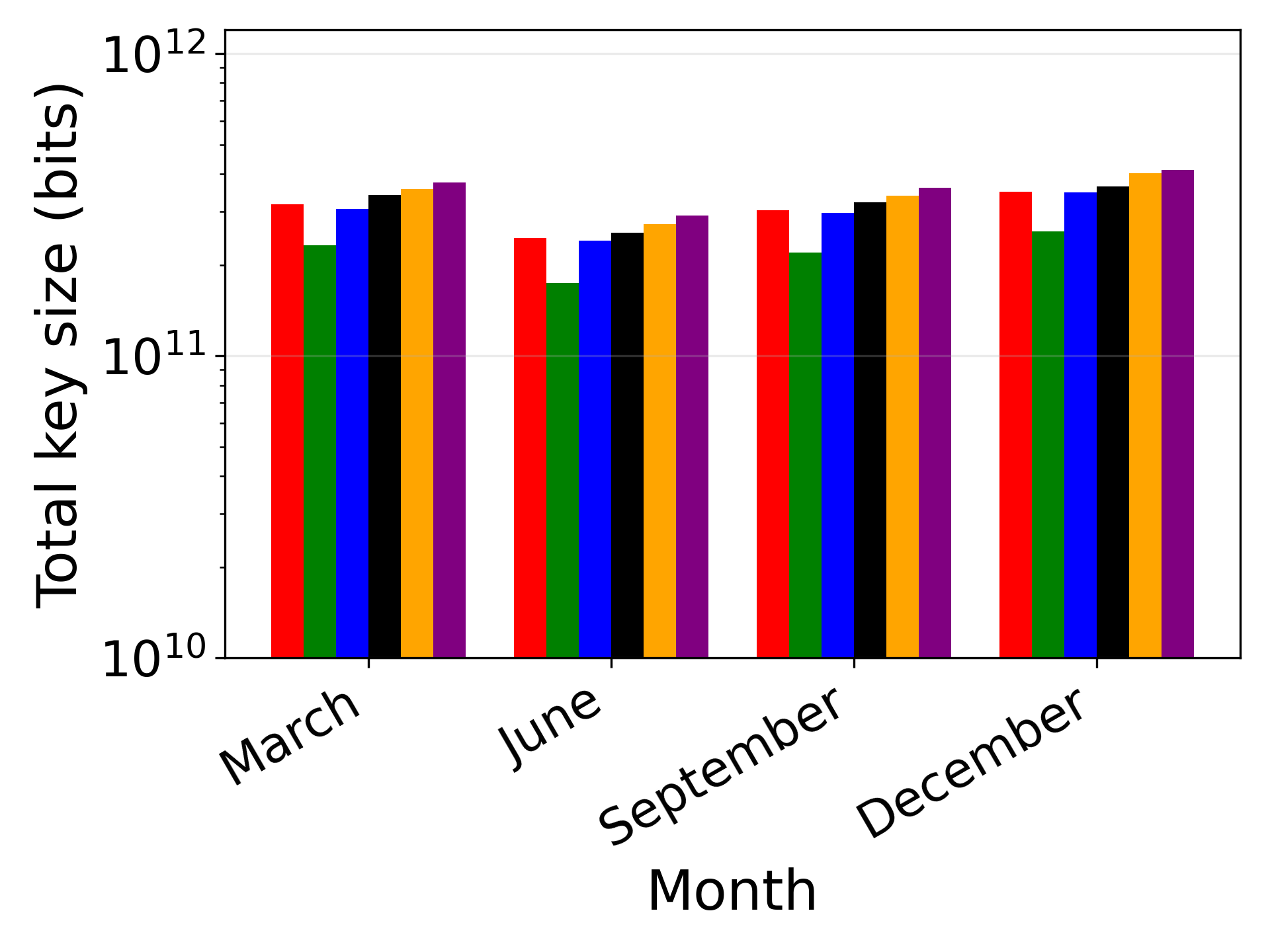}
        \caption{Assuming no cloud coverage.}
        \label{fig:totalkey_4gs_500_WOCC}
    \end{subfigure}
    %with cloud, no filtering
    % ---- Subfigure (a) ----
    \begin{subfigure}[b]{0.32\textwidth}
        \centering
        \includegraphics[width=\textwidth]{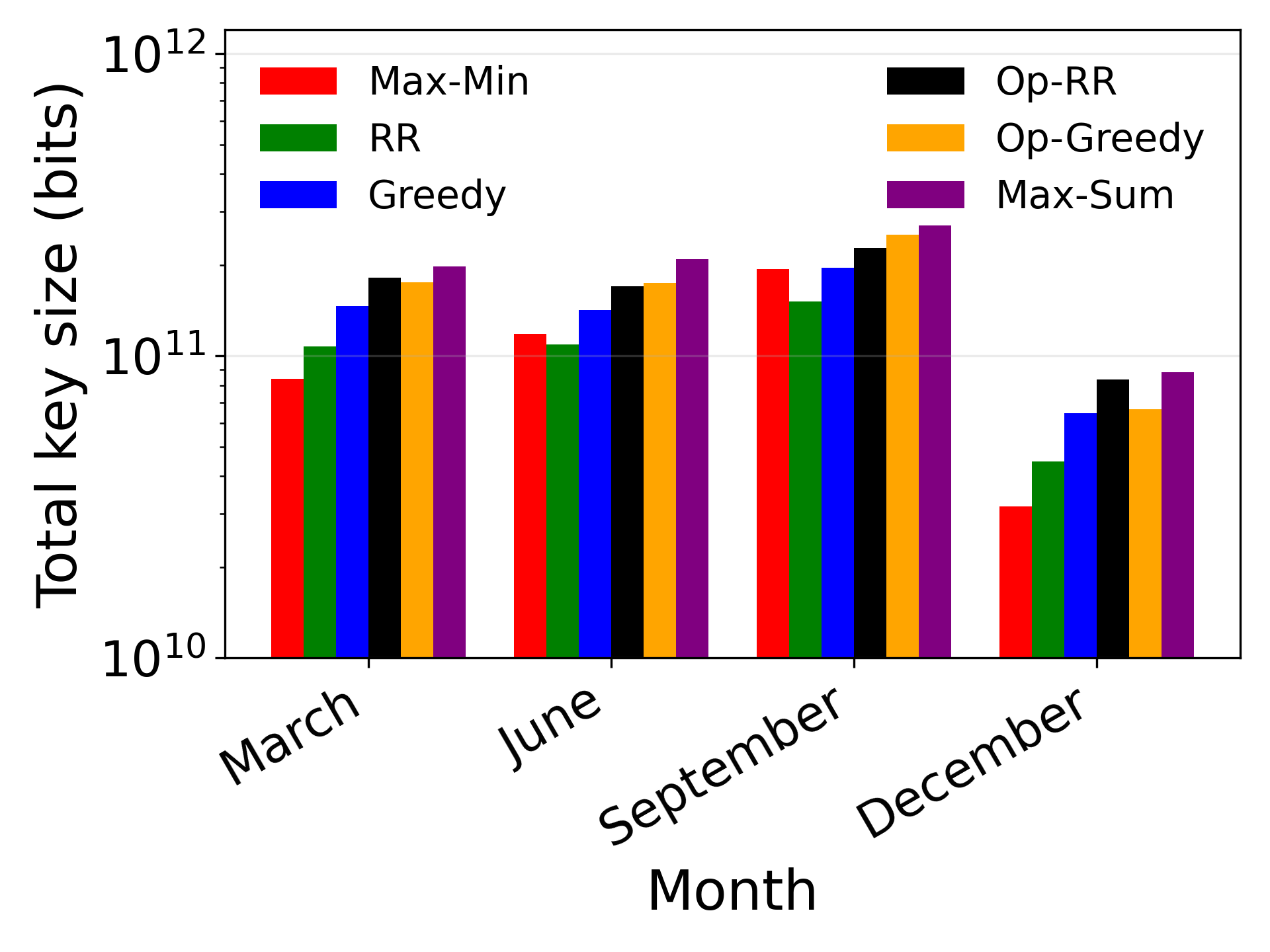}
        \caption{W/ cloud coverage, no filtering.}
        \label{fig:totalkey_4gs_500_WCC}
    \end{subfigure}
    %with cloud, with filtering
    \begin{subfigure}[b]{0.32\textwidth}
        \centering
        \includegraphics[width=\textwidth]{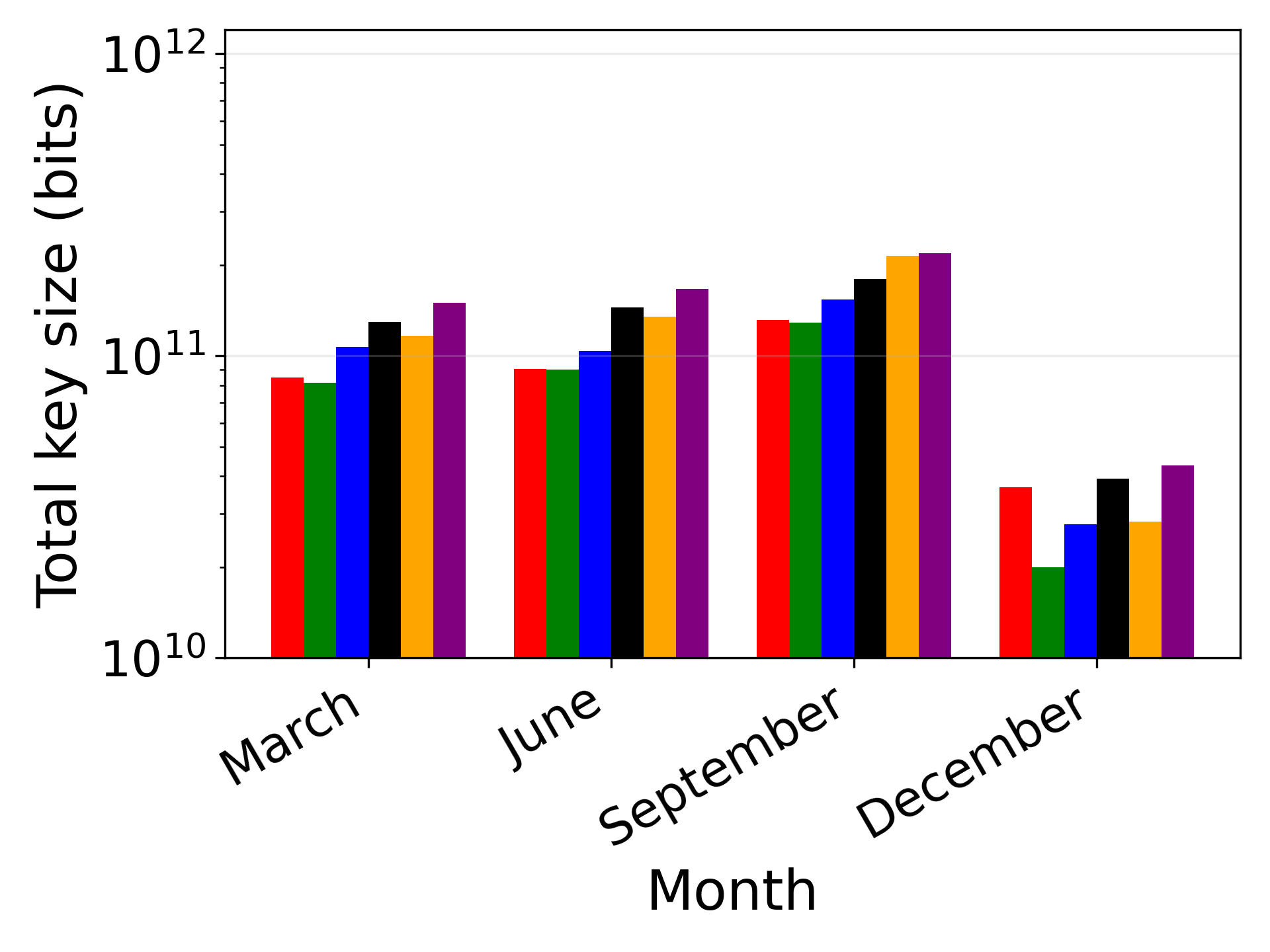}
        \caption{W/ cloud coverage, w/ filtering.}
        \label{fig:totalkey_4gs_500_CF}
    \end{subfigure}
    \caption{Total key size (bits) for the regional QKD setting, $A=500$ km. 
    }
    \label{fig:totalkey_4gs-500km}
\end{figure*}

%% file: related.tex
\section{Related Work} \label{sec:related}
%scheduling problems and solutions

In classical wireless networks, opportunistic scheduling refers to a class of popular scheduling policies that dynamically allocates resources to users with favorable channel conditions to improve overall resource utilization \cite{Asadi2013:opp-survey}. Most opportunistic scheduling frameworks \cite{andrews2007scheduling, liu2010scheduling, tassiulas2002scheduling, neely2003dynamic, khalil2011optimal} focus on maximizing network capacity, often neglecting fairness, which can disadvantage users with persistently weak channels. To address this, several studies such as \cite{Liu2001:oppor, Liu2003:oppor}, introduce fairness-aware scheduling policies with minimum-performance guarantees (MPG). In our work, we adapt the opportunistic scheduling principle to satellite-assisted QKD systems, building on the minimum-performance guarantee concept but extending it to a multi-satellite, multi-ground-station setting with visibility and hardware constraints, conditions not captured in classical MPG formulations.

%Our work adopts an opportunistic scheduling approach inspired by the minimum-performance guarantee framework to exploit the stochastic nature of satellite–to–ground quantum links. Unlike conventional wireless networks, our setup involves scheduling multiple satellites and ground stations under hardware constraints on transmitters and receivers.

Our scheduling problem differs fundamentally from the well-studied classical data transfer scheduling problem of satellite networks~\cite{Karapetyan15:downlink,Xhafa12:Genetic,Spangelo15:scheduling,Han18:Scheduling}, where satellites offload classical data to ground stations. They treat this problem using traditional resource allocation techniques such as Multiprocessor Scheduling, Resource-Constrained Project Scheduling, Genetic Algorithms, MIP programming. Those models involve only classical bits and do not account for the quantum or classical stages of QKD or the generation of shared secret keys among ground station pairs. 

The scheduling problem for dual-downlink entanglement distribution \cite{panigrahy2022optimal, williams2024scalable, chang2023entanglement, wei2024optimizing} and entanglement-based QKD \cite{Maule2024:scheduling} in satellite-based quantum networks  has been studied in prior work. However, these studies primarily focus on maintaining simultaneous connectivity between a ground-station pair and a satellite, and do not account for the opportunistic selection step that arises in the single-downlink scenario, where even a single ground station can be served at any point of time. The authors in~\cite{Polnik20:scheduling} consider single-downlink satellite QKD scheduling. 
They present an MIP  
formulation to schedule a single satellite to 
a set of 
ground stations with the goal of maximizing the sum of the key rates from the satellite to the ground stations. Our formulation differs significantly from that of \cite{Polnik20:scheduling}, since we consider generating key bits for ground station pairs assisted by 
multiple satellites, taking account of both total key bits and fairness. We present MIP formulations as comparison baselines, and opportunistic scheduling approaches.
%an opportunistic scheduling framework and solutions. 
In addition, our evaluation considers both global and regional QKD settings,
while the work in \cite{Polnik20:scheduling} focuses on a setting with ground stations all in the UK.   
%
%However, it considers a single satellite, while our approach generalizes to a constellation of satellites. 

%but it models the scheduling process analogous to classical data transfer.

%In our work, we employ opportunistic scheduling to exploit the stochastic nature of satellite–to–ground quantum links to improve secret key generation rates while maintaining fairness among ground stations over time.
%while providing fairness to users  \cite{Liu2001:oppor, Liu2003:oppor}

%This scheduling problem differs from the well-studied problem~\cite{Karapetyan15:downlink,Xhafa12:Genetic,Spangelo15:scheduling,Han18:Scheduling} of scheduling data transfer collected by satellites to ground stations in important aspects. The data transfer problem only involves  classical bits; it does not involve the quantum or classical stages of QKD, or the process of obtaining shared keys among ground stations.  % The data transfer problem does not involve QKD between a satellite and a ground station (neither the quantum or classical post-processing stage). In addition, it does not include the process of obtaining shared key bits among ground stations. 
%We are only aware of one study~\cite{Polnik20:scheduling} in this direction, which, however, treats QKD in the same way as data transfer. 

%% file: concl.tex
%\section{Conclusion and Future Work}

\section{Conclusion}

In this paper, we have developed an opportunistic scheduling framework for satellite-assisted QKD systems in single-downlink settings. This framework takes advantage of the dynamic, and diverse, satellite to ground station channels,  for efficient key establishment among ground station pairs assisted by the satellites. Under this framework, we developed two opportunistic scheduling schemes, Op-RR and Op-Greedy.    Using extensive simulation in a wide range of settings, we show that Op-RR achieves the best tradeoffs in fairness and total key size across all the schemes that we evaluate. We further show the impact of seasonal effects, cloud coverage, and the settings of the ground stations on the key establishment results, highlighting the importance of considering these realistic factors in evaluating satellite-assisted QKD systems.